\documentclass[12pt]{article}
\usepackage{cite,graphicx,amsmath,amssymb}
\usepackage{graphicx}

\def\lsim{\mbox{\raisebox{-.5mm}{$\,\stackrel{<}{\scriptstyle\sim}\,$}}}

\newcommand{\be}{\begin{equation}}
\newcommand{\ee}{\end{equation}}
\newcommand{\bea}{\begin{eqnarray}}
\newcommand{\eea}{\end{eqnarray}}
\newcommand{\uv}{\text{\tiny UV}}
\newcommand{\ir}{\text{\tiny IR}}

\newcommand{\cft}{\text{\tiny CFT}}

\newcommand{\sm}{\text{\tiny SM}}

\newcommand{\dec}{\text{\tiny dec}}
\newcommand{\rh}{\text{\tiny RH}}
\newcommand{\tot}{\text{\tiny tot}}

\begin{document}

\thispagestyle{empty}
\vspace*{.2cm}
\noindent
HD-THEP-09-26 \hfill 10 December 2009
\\
MPP-2009-200

\vspace*{1.0cm}

\begin{center}
{\Large\bf Tachyons in Throat Cosmology}
\\[1.5cm]

{\large S.~Halter$^{\,a}$, B.~v.~Harling$^{\,b}$, A.~Hebecker$^{\,c}$}\\[6mm]

{\it
$^{a}$~Max-Planck-Institut f\"ur Physik
(Werner-Heisenberg-Institut), F\"ohringer Ring 6, D-80805 M\"unchen,
Germany\\[3mm]
$^{b}$~School of Physics, University of Melbourne, Victoria 3010, 
Australia\\[3mm]
$^{c}$ Institut f\"ur Theoretische Physik, Universit\"at Heidelberg, 
Philosophenweg 16 und 19, D-69120 Heidelberg, Germany}
\\[.5cm]
{\small\tt (\,halter@mppmu.mpg.de} {\small ,} {\small\tt 
\,bvo@unimelb.edu.au} $\,\,${\small and} {\small\tt
a.hebecker@thphys.uni-heidelberg.de)}
\\[2.0cm]

{\bf Abstract}
\end{center} 

Tachyonic 5d scalars are generically present in Randall-Sundrum-like 
models. In particular, they are known to be part of the 5d effective 
description of the Klebanov-Strassler throat. When moving from the IR to 
the UV region, the 5d bulk profile of Kaluza-Klein excitations of tachyons decays more 
slowly than that of massless scalars or the graviton. As a result, tachyons 
in many cases dominate the coupling between IR- and UV-localized sectors, 
leading to a very significant enhancement of energy-transfer or decay rates
from the IR to the UV. This can dramatically affect the reheating of the 
Standard Model after brane inflation and the decay of throat dark matter.

\newpage
\section{Introduction}\label{intro}

Strongly warped regions or `throats' are a generic feature of type IIB 
flux compactifications~\cite{Giddings:2001yu,Denef:2004ze}. It can even be 
argued that, in certain regions of the `landscape', they are statistically 
unavoidable~\cite{Hebecker:2006bn}. The prime example is the
Klebanov-Strassler (KS) throat (or warped deformed conifold)
\cite{Klebanov:2000hb}. It can be viewed as a stringy version of a
Goldberger-Wise-stabilized Randall-Sundrum (RS) model~\cite{Brummer:2005sh} 
and is thus phenomenologically interesting for all the well-known reasons
making the RS model so attractive. Moreover, throats play a key role 
in mechanisms realizing de Sitter vacua \cite{Kachru:2003aw} and in 
string-theoretic models of inflation, such as brane-antibrane 
inflation \cite{Kachru:2003sx}.

In warped compactifications, couplings between fields localized in the UV and 
IR regions (regions of weak and strong warping, respectively) are 
generically suppressed by powers of the warp factor \cite{Dimopoulos:2001ui}. 
In cosmology, two implications of this suppression are of particular 
interest:$\,$\footnote{In 
non-cosmological applications of warped geometries, such as supersymmetry 
breaking and flavor model building, related effects are known as 
`sequestering' (see e.g. \cite{Randall:1998uk}). The technical questions
arising in this context are somewhat different and we will not address 
them in the following.
}
One is the long lifetime of certain IR-localized modes, which can hence play 
the role of `throat dark matter' \cite{Chen:2006ni,Berndsen:2008my,
Dufaux:2008br,Harling:2008px,Chen:2009iua,Frey:2009qb}. The other is the small energy 
transfer rate from a plasma of such modes (created at the end of warped 
brane inflation) to any UV-localized sector \cite{Barnaby:2004gg,
Kofman:2005yz,Chialva:2005zy,Frey:2005jk,Panda:2009ji}. This can cause problems for the 
reheating of the Standard model after inflation in a throat. 

The main point of the present paper is to demonstrate that tachyonic 5d 
scalars can play a central role in coupling IR- and UV-localized sectors 
and to discuss some of the immediate cosmological implications. 

The importance of tachyons in this context is easily understood: 
Recall first that the presence of tachyons is AdS$_5$ is natural from the 
perspective of 5d supersymmetry \cite{Shuster:1999zf,Gherghetta:2000qt} 
and observed in concrete examples, such as the compactification of type IIB 
supergravity to AdS$_5$ on $T^{1,1}$ \cite{Ceresole:1999zs}. It is 
well-known that tachyons do not destabilize infinite AdS$_5$ as long as 
their negative mass-squared respects the Breitenlohner-Freedman bound 
\cite{Breitenlohner:1982jf}. On a slice of AdS$_5$, tachyons generically 
introduce an instability. In supersymmetric models on a slice of AdS$_5$, 
stability is maintained by a positive mass-squared operator localized at 
the UV brane \cite{Gherghetta:2000qt}. Such a brane-mass term can also be 
used to ensure the stability of non-supersymmetric Randall-Sundrum models 
with tachyons \cite{Ghoroku:2001pi,Delgado:2003tx,Kaplan:2009kr}.\footnote{Alternatively, 
one can impose the Dirichlet boundary condition at the UV brane \cite{Toharia:2008ug}. 
This corresponds to the limit of sending the boundary mass to infinity.} Now, for 
any massive 5d scalar, its efficiency in mediating interactions between IR 
and UV brane is determined by the 5d bulk profile of its lowest Kaluza-Klein 
modes. Generically, the value of IR-localized modes is exponentially 
suppressed near the UV brane. This exponential suppression is governed by 
the 5d mass-squared. Thus, it is natural to expect that {\it 
tachyonic} scalars will provide the strongest coupling between the UV 
and IR brane. In the following, we study this general idea at the 
quantitative level. 

Our main point can be made in a simple 5d toy model: a single 
tachyonic scalar on a slice of AdS$_5$. Any of its low-lying IR-localized 
KK modes can be taken to represent a typical particle of 
the IR sector. The UV sector to which such a KK mode might 
decay is modeled as a 4d gauge theory localized on the UV brane. Our 5d
scalar couples to the $F^2$ term of this theory through its value 
at the UV-brane. We calculate the corresponding decay rate and find it to
be of the same order of magnitude as that resulting from the mediation of 
the graviton.\footnote{The 
bulk profile of modes of the 5d graviton is the same as that of a 
massless 5d scalar. 
}
This result arises from two compensating effects: On the one hand, the 
modes of a tachyon decay more slowly than those of a graviton when moving 
from the IR to the UV. On the other hand, they are subject to an extra 
suppression due to the UV-localized mass term (which is necessary for 
stability). 

From the above, it is clear that in many relevant situations the tachyon 
is, in fact, bound to dominate: Namely, massless 5d scalars are expected 
to obtain a large UV mass term in generic, fully stabilized models. The 5d
graviton, which has the profile of a massless scalar and no UV mass term, 
can not mediate the decay of spin-zero states. Thus, the tachyon dominates
the decay of low-lying IR-localized spin-zero states or `glueballs'. 
We illustrate this in a simplified KKLMMT-type setup for warped reheating. 
It turns out that tachyons dominate the energy transfer to the Standard 
Model. They can thus help to avoid the complete dissipation of inflationary 
energy density into gravitons, thereby solving a generic problem of (warped) 
brane inflation. 

Our paper is organized as follows: Section~\ref{AdS5fields} provides a 
pedagogical introduction to tachyons in a slice of AdS$_5$. Building on this 
discussion, Sect.~\ref{adsdecays} gives a simple and intuitive derivation 
of the decay rate of the IR-localized modes of a 5d tachyon to a 
UV-localized sector. It is followed by a discussion of situations in 
which the tachyon dominates the decay of IR-localized states in 
Sect.~\ref{tachyondominated}. In Section~\ref{cftpicture}, we explain our 
results from the perspective of the dual CFT, finding further support for 
our intuitive picture and our calculations. We then discuss the applicability 
of our 5d analysis to presumably more realistic and general throat geometries 
in Sect.~\ref{ApplicationstoThroats}. Finally, in Sect.~\ref{cosmology}, 
we turn to cosmological applications, in particular to effects 
on reheating and dark matter decay. Our Conclusions are followed by an
Appendix in which we provide a Bessel function analysis supporting our 
order-of-magnitude calculation in the main text.

Cosmological implications of tachyonic 5d scalars in a warped throat were
discussed in \cite{Berndsen:2008my} and, more recently, in
\cite{Frey:2009qb,Sundrum:2009ii}. The analysis of \cite{Frey:2009qb}
includes processes which, in our language, can be interpreted as
tachyon-mediated decays from the IR- to the UV-localized sector. We
discuss the relation to our work in Sect.~\ref{ApplicationstoThroats}.
In \cite{Sundrum:2009ii}, the characteristic 5d profile of tachyons is
used in the construction of the waterfall-sector of a model of hybrid
inflation.

\section{Tachyons in a slice of AdS$_{5}$}
\label{AdS5fields}
To set up our notation, we recall that an RS I model \cite{Randall:1999ee} 
is defined by a slice of AdS$_{5}$ with metric
\begin{equation}
\label{AdS5metric}
ds^2 = e^{-2 k y}(\eta_{\mu \nu} dx^{\mu} dx^{\nu}) + dy^2 \, ,
\end{equation}
where $k = 1/R$ is the inverse curvature radius. The slice is bounded by 
two 3-branes at $y_{\uv}=0$ and $y_{\ir}=\ell$. In the following, we will
focus on the dynamics of a scalar $\Phi$ with action 
\begin{equation}
\label{5daction}
S_{5d} \, = \, \int d^{4}x dy\,\sqrt{-G}\,\frac{1}{2}\,\Phi \left[
\nabla^2 - M^{2}  \right] \Phi\,,
\end{equation}
where $\nabla^2 = (\sqrt{-G})^{-1} \partial_M (\sqrt{-G} G^{MN} 
\partial_N)$ and $G_{MN}$ is the AdS$_5$ metric. Capital indices run 
over $\{ 0, \ldots, 3\}$ and $y$. We assume that the model is already 
stabilized, e.g. via the Goldberger-Wise mechanism \cite{Goldberger:1999uk}, 
and that $\Phi$ is an additional scalar field, not related to the 
stabilization.

It will be convenient to use the coordinate $z=k^{-1}e^{k y}$ and the 
corresponding conformally flat metric 
\begin{equation}
\label{conformalAdS5metric}
ds^2 \, = \, \frac{1}{(k z)^2}( \eta_{\mu \nu} dx^{\mu} dx^{\nu} + dz^2 ) \, .
\end{equation}
Expanding $\Phi$ in KK modes,
\begin{equation}
\label{KKdecomposition}
\Phi(x,z) \,=\, \sum_{n} \varphi_{n}(z) \chi_{n}(x) \, ,
\end{equation}
we have a set of canonically normalized 4d scalars $\chi_n$ with 
masses $m_n$, the wave functions of which satisfy the equations 
\begin{equation}
\label{5dprofileequation}
\left( z^{2} \partial_{z}^{2} - 3 z \partial_{z} + m_{n}^{2} z^{2} -
\frac{M^{2}}{k^{2}} \right) \varphi_{n}(z) \, = \, 0 \, .
\end{equation}
To specify boundary conditions, it is convenient to view the extra dimension 
as an $S^{1}/\mathbb{Z}_{2}$ orbifold. Thus, our $\varphi_n$ are defined 
on the double cover of the interval $[z_{\uv},z_{\ir}]$, where $z_{\uv}= 
k^{-1}$ and $z_{\ir}=k^{-1} e^{k \ell}$. We assume that $\Phi$ is even under 
the $\mathbb{Z}_2$ symmetry transformation, implying that $\varphi_n$ is 
symmetric under $(z-z_{\uv})\to -(z-z_{\uv})$ and periodic with period 
$2(z_{\ir}-z_{\uv})$.

In terms of the rescaled field 
\be
\label{rf}
\psi_{n}(z) \equiv (zk)^{-3/2} \varphi_{n}(z)\,,
\ee
Equation~\eqref{5dprofileequation} takes the form of a 1-dimensional 
Schr\"odinger equation
\cite{Randall:1999ee}
\begin{equation}
\label{schroedinger}
\left( - \partial_{z}^{2} + V(z) \right) \psi_{n}(z) \, = \, m_{n}^{2} \,
\psi_{n}(z)
\, 
\end{equation}
with `energy' $m_{n}^{2}$ and potential
\begin{equation}
\label{potentialz}
V(z) \, = \, \frac{\alpha^{2} - \frac{1}{4}}{z^{2}} - \frac{3}{z_{\uv}}
\delta(z - z_{\uv}) + \frac{3}{z_{\ir}} \delta(z - z_{\ir}) \,.
\end{equation}
Here $\alpha \equiv \sqrt{4 + M^{2}/k^{2}}\,$, and the $\delta$-function 
contributions come from the rescaling in the presence of boundary 
conditions. The solutions of Eq.~\eqref{schroedinger} are Bessel functions 
of order $\alpha$ (cf.~Appendix~\ref{BesselCalc}) and all that follows 
could be derived from a careful analysis of the behaviour of these 
functions. However, we find it more illuminating to make our main 
qualitative points using a parametric analysis of approximate but explicit 
solutions. 

\begin{figure}[ht]
\begin{center}
\includegraphics[scale=0.4]{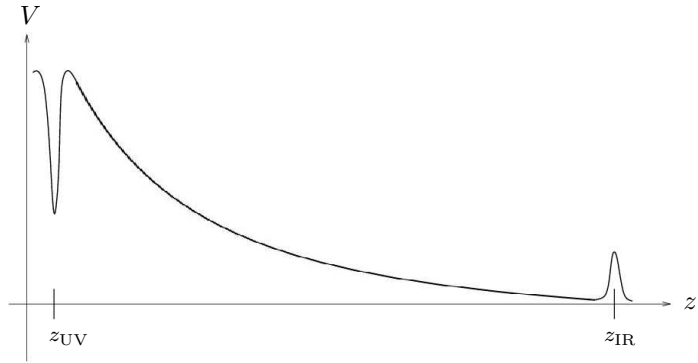}
\put(-247,128){\footnotesize $V$}
\put(4,20){\footnotesize $z$}
\put(-26,8){\scriptsize $z_{\ir}$}
\put(-236,8){\scriptsize $z_{\uv}$}
\caption{Potential in the effective Schr\"o\-din\-ger equation for a 5d mass
$M = 0$.}
\label{potmassless}
\end{center}
\end{figure}

\begin{figure}[ht]
\begin{center}
\includegraphics[scale=0.4]{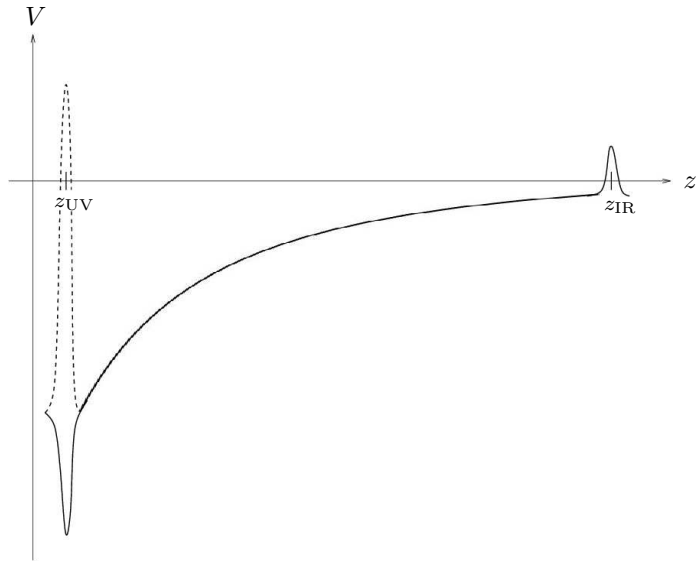}
\put(-245,203){\footnotesize $V$}
\put(4,141){\footnotesize $z$}
\put(-26,133){\scriptsize $z_{\ir}$}
\put(-234,134){\scriptsize $z_{\uv}$}
\caption{Potential in the effective Schr\"o\-din\-ger equation for a
5d mass $M^2 =-4 k^2$. The dotted $\delta$-peak appears when
a large mass term on the UV brane is added.}
\label{potTachyon}
\end{center}
\end{figure}

According to Eqs.~\eqref{schroedinger} and \eqref{potentialz}, we are 
basically looking for the solutions of a quantum mechanical problem on an
interval.\footnote{More 
precisely, we are looking for the even solutions on the $S^1$ covering 
space of that interval.
}
The coefficient of the $1/z^2$ term in the potential 
is subject to the Breitenlohner-Freedman bound 
\cite{Breitenlohner:1982jf} $M^2 \geq - 4 k^2$ (corresponding to 
$\alpha \geq 0$), which ensures the stability of the AdS$_5$ vacuum. 
Figures~\ref{potmassless} and~\ref{potTachyon} show the potential for the 
particularly interesting cases $M = 0$ and $M=-4k^2$. Standard KK-mode 
intuition tells us that all $m_n^2$ are positive for $M^2>0$, while 
$m_0=0$ for $M=0$. We then clearly expect the presence of 
negative `energy eigenvalues' $m_n^2$ for `tachyonic' $M^2<0$. This is not 
tolerable from the perspective of the 4d effective field theory. Indeed, 
while the Breitenlohner-Freedman bound is sufficient to ensure the 
stability of infinite AdS space, it does not guarantee the stability of a 
slice of AdS. 

One way to avoid this instability is to introduce a UV-brane-localized 
mass operator\cite{Ghoroku:2001pi,Delgado:2003tx,Kaplan:2009kr}. We 
therefore supplement Eq.~(\ref{5daction}) with
\begin{equation}
\label{branemassterms}
S_{\uv} \, = \, - \int d^{4}x dy \sqrt{-G} \, \lambda k \, \Phi^{2}(x,y)
\delta(y - y_{\uv})\,,
\end{equation}
where we have written the dimensionful coefficient as a product of $k$
and a dimensionless parameter $\lambda \geq 0$. This mass term modifies 
the potential, 
\begin{equation}
\label{fullpotential}
V(z) \, = \, \frac{\alpha^{2} - \frac{1}{4}}{z^{2}} + \frac{2 \lambda -
3}{z_{\uv}} \delta(z - z_{\uv}) + \frac{3}{z_{\ir}} \delta(z - z_{\ir}) \,,
\end{equation}
such that, for $\lambda>\frac{3}{2}$, the $\delta$-peak at the UV brane 
becomes repulsive (cf.~Fig.~\ref{potTachyon}). It is intuitively clear
that such a modification of the potential lifts the energy eigenvalues 
and can therefore remove tachyonic modes. 

Indeed, it will become apparent in the next section that, in the limit 
$z_{\ir} \rightarrow \infty$ (the RS II limit \cite{Randall:1999vf}), a 
zero mode appears if $\lambda=\lambda_0\equiv 2-\alpha$. This is the 
minimal value of $\lambda$ required for stability -- for 
$\lambda>\lambda_0$ all modes are massive. In the RS I case (for finite 
$z_{\ir}$) the minimal value of $\lambda$ may differ from $\lambda_0$, 
depending on the specific boundary condition imposed at the IR brane. 
However, this difference goes to zero together with $z_{\uv}/z_{\ir}$ and 
its precise value will not be important. Moreover, for the specific case 
of supersymmetric boundary conditions (when the IR brane carries as mass 
term of opposite value w.r.t. the UV-brane mass term~\cite{
Gherghetta:2000qt}), the minimal value continues to be $\lambda_0$. Thus, 
it will be convenient for us to assume such supersymmetric IR boundary 
conditions (modifying Eq.~(\ref{fullpotential}) appropriately) and to think 
of $\lambda_0=2-\alpha$ as of the minimal value of $\lambda$, also at 
finite $z_{\ir}$. 

To conclude our preliminary discussion of tachyons in a slice of AdS$_5$,
we briefly describe the KK mass spectrum. As long as $\lambda>\lambda_0$, 
the spectrum does not differ qualitatively from the familiar massless or
positive-mass-squared case. As we already emphasized, we are dealing with 
a quantum-mechanical problem on a compact space of size $\sim 
(z_{\ir}-z_{\uv}) \simeq z_{\ir}$. We thus naively$\,$\footnote{This 
is naive in the sense that the potential introduces the further 
dimensionful quantity $z_{\uv}$ into the calculation. However, in the 
limit $z_{\uv}/z_{\ir} \to 0$ the $1/z^2$ term together with the 
UV-brane $\delta$ function basically just serve to provide a certain 
boundary condition at one side of the interval. Hence the scale $z_{\uv}$ 
does not affect the (low-lying part of) the spectrum. This will become 
apparent after the analysis of the next section.
} 
expect $m_n$ to be quantized in units of $z_{\ir}^{-1}$, to which we will 
from now on refer as our {\it IR scale} $m_{\ir}\equiv z{\ir}^{-1} = k\,
e^{-k\ell}$. Indeed, the Bessel function analysis of 
App.~\ref{BesselCalc} shows that $m_n \simeq (n+\frac{\alpha}{2}-
\frac{3}{4})\,\pi\, k \, e^{- k \ell}$ for $1\ll n\ll e^{k\ell}$. This 
approximate formula remains rather accurate all the way down to relatively 
small $n$. 

We finally note that a technically related analysis of the wave functions 
of a tachyonic scalar has recently appeared in~\cite{Kaplan:2009kr} in 
a rather different physical context. We refer to that paper for more 
details on the quantum mechanical analogue and corresponding references.

\section{Decays mediated by AdS$_5$ fields}
\label{adsdecays}
As one can see from Fig.~\ref{potmassless}, a massless scalar has to 
tunnel through an effective potential barrier before it can reach the UV 
brane.\footnote{This is the same situation as for the graviton 
(cf.~\cite{Randall:1999vf}).} By contrast, a tachyonic scalar with $M^2= 
-4 k^2$ has no such barrier to  surmount (cf.~Fig.~\ref{potTachyon}). 
Thus, one expects the tachyon to couple more strongly to the UV brane.
There is, however, a compensating effect: The UV-brane mass term of the 
tachyon suppresses its value at the brane. To determine the relative 
importance of these two effects, we now estimate the decay rate of 
the KK modes of a bulk scalar to a gauge theory living on the UV brane. 
Our interest is in the dependence on the parameters $\alpha$ and $\lambda$.

\subsection{Generic UV-brane mass term}
We assume that $\Phi$ interacts with a gauge theory localized at the UV 
brane via the action\footnote{As we will discuss in 
Sect.~\ref{ApplicationstoThroats} in more detail, this term can be viewed 
as an effective 5d description of the dilaton coupling to the gauge fields 
on a D-brane stack.} \begin{equation} \label{assumedcoupling} \int d^{4}x 
\, dz \, \sqrt{-G} \,k^{-\frac{3}{2}} \, \Phi(x,z)\, \mbox{tr} \left( 
F^{\mu\nu} F_{\mu \nu}\right) \, \delta(z - z_{\uv}) \, . \end{equation} 
We have arbitrarily set the coefficient of this interaction term to one in 
units of $k$ since we will anyway perform only an order-of-magnitude 
calculation. Using Eq.~\eqref{KKdecomposition}, we see that the coupling 
of the $n$-th KK mode to the gauge theory is given by \begin{equation} 
\label{KKcoupling} \int d^{4}x \,k^{-\frac{3}{2}} \,\varphi_{n}(z_{\uv}) 
\, \mbox{tr}\left(F^{\mu \nu} F_{\mu \nu}\right) \, \chi_n(x) \, . 
\end{equation} The coupling constant between the $n$-th KK mode and two 
gauge bosons,
\begin{equation}
\label{dimfulcouplingUV}
g_n \sim k^{-\frac{3}{2}}\,\varphi_{n}(z_{\uv})=k^{-\frac{3}{2}}\,
\psi_n(z_{\uv})\,,
\end{equation}
has mass dimension $[g_n]=-1$ and determines the decay rate according to
\begin{equation}
\label{decayrate}
\Gamma_n \, \sim \, \displaystyle g_{n}^{2} \, m_{n}^{3} \, .
\end{equation}

In order to calculate $g_n$, we need a detailed understanding of the shape of 
the KK modes. To achieve this, we introduce the coordinate $\widehat{z}=m_n z$ 
(for each $m_n \neq 0$) and translate the $\delta$ functions back to boundary 
conditions at $\widehat{z}_{\uv}$ and $\widehat{z}_{\ir}$. We then have to 
solve
\begin{equation}
\label{schroedinger2}
\left[ \partial_{\widehat{z}}^{2} + 1 - \left(\frac{\alpha^{2} -
\frac{1}{4}}{\widehat{z}^{2}}\right) \right] \psi_{n}(\widehat{z}) \, = \, 0
\, ,
\end{equation}
subject to
\begin{equation}
\label{fullBCz}
\displaystyle \bigl(\widehat{z} \,  \partial_{\widehat{z}}
\psi_{n}\bigr)\vert_{\widehat{z} = \widehat{z}_{\uv}}\, = \, \left(\lambda -
\frac{3}{2}\right) \,
\psi_{n} \vert_{\widehat{z} = \widehat{z}_{\uv}} \quad \text{and} \quad
\bigl(\widehat{z} \, 
\partial_{\widehat{z}}\psi_{n}\bigr)\vert_{\widehat{z} = \widehat{z}_{\ir}}\,
= \,\left(\lambda-\frac{3}{2}\right) \, \psi_{n} \vert_{\widehat{z} = 
\widehat{z}_{\ir}} \,,
\end{equation}
where we have also modified the IR boundary condition as mentioned 
earlier. 

\begin{figure}
\centering
\includegraphics[scale=0.4]{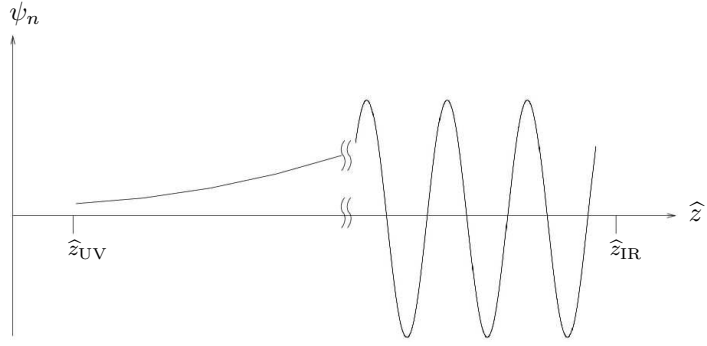}
\put(-253,122){\footnotesize $\psi_n$}
\put(4,46){\footnotesize $\widehat{z}$}
\put(-26,33){\scriptsize $\widehat{z}_{\ir}$}
\put(-231,33){\scriptsize $\widehat{z}_{\uv}$}
\caption{Sketch of the wavefunction $\psi_{n}$ as a function of
$\widehat{z}$}
\label{sketch}
\end{figure}

It is straightforward to solve Eq.~\eqref{schroedinger2} separately in the 
regions $\widehat{z}\ll 1$ and $\widehat{z}\gg 1$, where the first and second 
term of the potential, respectively, can be neglected.\footnote{We
assume $\alpha^2-\frac{1}{4}\sim {\cal O}(1)$ for simplicity, excluding 
the special region $\alpha\simeq \frac{1}{2}$. 
}
We find that for small $\widehat{z}$
\begin{equation}
\label{smallzsol}
\psi_{n, \uv}(\widehat{z}) \, \simeq \, \begin{cases}
\displaystyle \frac{1}{N_{\alpha}} \left(\widehat{z}^{\frac{1}{2} + \alpha} +
B_{\alpha} \, \widehat{z}^{\frac{1}{2} - \alpha}\right) & \text{for} \, 
\alpha > 0 \\
 & \\
\displaystyle \frac{1}{N_{0}} \left( \widehat{z}^{\frac{1}{2}} + B_{0} \,
\widehat{z}^{\frac{1}{2}} \ln \frac{1}{\widehat{z}} \right) & \text{for} 
\, \alpha = 0 \,,
\end{cases}
\end{equation}
while for large $\widehat{z}$
\begin{equation}
\label{largez}
\psi_{n, \ir}(\widehat{z}) \, \simeq \, \frac{1}{A_{\alpha}}
\cos(\widehat{z} + C_\alpha) \,,
\end{equation}
where $N_\alpha$, $B_\alpha$, $A_\alpha$ and $C_\alpha$ are 
constants of integration.\footnote{Note 
that, with the redefinitions 
\begin{equation*}
\label{redefinitionforalphato0}
N_0 \equiv \frac{N_{\alpha}}{1 + B_{\alpha}} \quad \text{and} \quad B_0 \equiv
\frac{\alpha (1 - B_{\alpha})}{1 + B_{\alpha}} \,,
\end{equation*}
the second line of Eq.~(\ref{smallzsol}) can be recovered as the 
$\alpha~\rightarrow~0$ limit of the first line.
}
The qualitative behaviour of the resulting wave function is sketched in 
Fig.~\ref{sketch}.

The canonical normalization of $\chi_n$ implies, together with 
Eq.~(\ref{rf}), the normalization condition 
\begin{equation}
\label{normalization}
\frac{1}{m_n} \int_{\widehat{z}_{\uv}}^{\widehat{z}_{\ir}}
d\widehat{z} \; \psi_{n}(\widehat{z})^2 \, = \, 1
\end{equation}
for our wave function. Let us assume that $\widehat{z}_{\ir}\gg 1$ and that 
the absolute value of the wave function grows between $\widehat{z}_{\uv}$ and 
$\widehat{z}\sim 1$ (this turns out to always be the case if $\lambda$ is 
generic). It is then apparent that the overall normalization is dominated by 
the cosine solution in the IR region $1\leq\widehat{z}\leq \widehat{z}_{\ir}$, 
implying
\begin{equation}
\label{approxN}
A_\alpha \sim \sqrt{\widehat{z}_{\ir}/m_n}=m_{\ir}^{-1/2}\, .
\end{equation}
Of course, we are actually interested in the lowest-lying modes, so that 
$z_{\ir}\sim {\cal O}(1)$ rather than $z_{\ir}\gg 1$. Nevertheless, the 
approximate cosine region (which in this case contains only a few 
oscillations) continues to contribute an ${\cal O}(1)$ fraction to the 
total normalization. Thus, Eq.~(\ref{approxN}) continues to be correct up 
to ${\cal O}(1)$ factors. 

To summarize, we now know that the UV solution has to be matched to 
the IR solution with approximate value $A_\alpha^{-1}\sim m_{\ir}^{1/2}$
at $\widehat{z}\sim 1$. For the decay rate, we need the UV-brane value of
this UV solution, which we will now derive.

The first boundary condition of Eq.~(\ref{fullBCz}) together with 
Eq.~(\ref{smallzsol}) gives
\bea 
\label{balphadetuned}
B_\alpha &=& -\left(1-\frac{\,2\alpha}{\lambda-2+\alpha}\right)
\widehat{z}_{\uv}^{\,2\alpha}\,\,\,,\label{ba}\\ \nonumber\\
B_0 &=& -\left(1-\frac{1}{(\lambda-2)\ln(1/\widehat{z}_{\uv})+1}\right)
\,\frac{1}{\ln(1/\widehat{z}_{\uv})}\,.\label{b0}
\eea
Now, for $\lambda=\lambda_0+{\cal O}(1)$ and $\alpha={\cal O}(1)$, we 
see that both the $\widehat{z}^{\,\frac{1}{2}+\alpha}$ solution and 
the $\widehat{z}^{\,\frac{1}{2}-\alpha}$ solution, as well as the 
full solution $\psi_{n,\uv}$ are of comparable size at $\widehat{z}_{\uv}$. 
Thus, only the more strongly growing solution is important for the 
matching at $\widehat{z}\sim 1$. The value of $\psi_n$ at the UV brane 
is suppressed according to the behaviour of this solution. By contrast, 
for $\alpha=0$ the $\widehat{z}^{\,\frac{1}{2}}$ solution and the 
$\widehat{z}^{\,\frac{1}{2}}\ln(1/\widehat{z})$ solution cancel almost 
exactly at the UV brane, leading to an extra suppression factor 
$1/\ln(1/\widehat{z})$ in the brane value of $\psi_n$. In summary, we
have
\be
\psi_n(\widehat{z}_{\uv}) \, \sim \, \begin{cases}
\displaystyle \widehat{z}_{\uv}^{\,\frac{1}{2}+
\alpha}m_{\ir}^{\frac{1}{2}} & \text{for} \, \alpha > 0 \\
 & \\
\displaystyle \frac{\widehat{z}_{\uv}^{\,\frac{1}{2}}}{\ln(1/
\widehat{z}_{\uv})}m_{\ir}^{\frac{1}{2}} & \text{for} \, \alpha = 0 \, .
\end{cases}
\ee
Note that the $\alpha=0$ result could also have been derived by using the 
expression for $B_\alpha$ in Eq.~(\ref{ba}) and carefully taking the 
limit $\alpha\to 0$. It is valid for $\alpha\lsim\ln(1/\widehat{z}_{\uv}
)^{-1}$. 

Using Eqs.~(\ref{dimfulcouplingUV}) and (\ref{decayrate}) and recalling  
that $\widehat{z}_{\uv}=m_n/k$, one finds the decay rates
\be
\label{drs}
\Gamma_n \, \sim \, \begin{cases}
\displaystyle \left(\frac{m_n}{k}\right)^{4+2\alpha}m_{\ir} & \text{for} 
\, \alpha > 0 \\
 & \\
\displaystyle 
\left(\frac{m_n}{k}\right)^4\frac{m_{\ir}}{\ln^2(k/m_n)} & \text{for} \, 
\alpha = 0 \, .
\end{cases}
\ee
These rates represent one of our main results and will be used in the 
cosmological applications later on.

\subsection{Tuned UV-brane mass term}\label{tmt}
Equation~(\ref{ba}) suggests that the point $\lambda=\lambda_0=2-\alpha$ 
may require special attention: Near this point, the UV-brane value of the 
$\widehat{z}^{\,\frac{1}{2}-\alpha}$ mode (the falling or weakly growing 
mode) is significantly enhanced compared to the value of the 
$\widehat{z}^{\,\frac{1}{2}+\alpha}$ mode (the strongly growing mode). We 
already know that, for $\lambda=\lambda_0$, the lowest mode is massless 
since, for the specific choice of IR boundary conditions discussed 
earlier, $\lambda_0$ is the minimal allowed value. This mode can not decay
unless the tuning is imperfect. We will not discuss the possible decay rate 
of such an `almost massless' mode. 

Furthermore, in the tuned case the UV-brane 
$\delta$-function is least-repulsive (or most attractive). Thus, we expect 
the decay rates of higher modes to be larger for $\lambda=\lambda_0$ than 
for the generic case of $\lambda>\lambda_0$. These enhanced decay rates are
our main interest in this subsection.

Let us first consider the case $0<\alpha<1$ and $\lambda=\lambda_0$.
Taking Eq.~(\ref{ba}) at face value, $B_\alpha$ is infinite, which simply 
means that in the solution of Eq.~(\ref{smallzsol}) only the 
$\widehat{z}^{\,\frac{1}{2}-\alpha}$ term is present. When moving from 
$\widehat{z}={\cal O}(1)$ to $\widehat{z}=\widehat{z}_{\uv}$, the 
wave function then changes by a factor $\smash{\widehat{z}_{\uv}^{\,
\frac{1}{2}-\alpha}}$. Repeating the arguments which lead us to 
Eq.~(\ref{drs}), this gives the decay rate
\be
\label{0a1}
\Gamma_n\sim \left(\frac{m_n}{k}\right)^{4-2\alpha}m_{\ir} \quad 
\text{for} \quad 0<\alpha <1 \quad \mbox{and} \quad \lambda=\lambda_0\,.
\ee

In the case $\alpha=0$, we have $B_0=0$ at $\lambda=\lambda_0$ according 
to Eq.~(\ref{b0}). Hence the wave function falls by a factor 
$\smash{\widehat{z}_\uv^{\,\frac{1}{2}}}$. This implies
\be
\label{a0}
\Gamma_n\sim \left(\frac{m_n}{k}\right)^4 m_{\ir} \quad \text{for} \quad
\alpha = 0 \quad \mbox{and} \quad \lambda=\lambda_0 \, .
\ee
As expected, a tuned value of $\lambda$ enhances the rates in the whole 
range $0\leq\alpha<1$, with the enhancement being more pronounced at 
larger $\alpha$. 

Apparently, nothing in the above calculation depends on the restriction to
$\alpha<1$. However, it is easy to see that something goes wrong for 
$\alpha> 1$. Indeed, if only the $\widehat{z}^{\,\frac{1}{2}-\alpha}$ 
mode is relevant for all $\widehat{z}\ll 1$, the $\psi_n^2$ 
normalization integral is dominated by the region near 
$\widehat{z}_{\uv}$. Thus, the relevant mode is UV localized. Obviously, 
this can not be the case for all modes with $m_n\ll k$ (to which
our argument formally applies).

At the technical level, the above problem comes from the insufficient 
accuracy of the approximation made in Eq.~(\ref{smallzsol}). This 
accuracy can be improved by taking into account higher-order terms in the 
small-argument expansions of the Bessel functions.\footnote{Alternatively, 
they can be determined by solving Eq.~(\ref{schroedinger2}) iteratively,
treating the `1' as a perturbation.
}
Including only the first-order correction, we find
\begin{equation}
\label{smallzcorrect}
\psi_{n, \uv}(\widehat{z}) \, \simeq \, \begin{cases}
\displaystyle \frac{1}{N_{\alpha}}
\left[\widehat{z}^{\,\frac{1}{2}+\alpha}
\left( 1 - \frac{\widehat{z}^2}{4 (1+\alpha)} \right) +
B_{\alpha} \, \widehat{z}^{\,\frac{1}{2}-\alpha} \left(1-\frac{\widehat{z}^2}{4
(1-\alpha)}\right)\right] & \text{for} \, \; \alpha \neq 1 \\
 & \\
\displaystyle \frac{1}{N_1}
\left[\widehat{z}^{\,\frac{3}{2}}
\left( 1 - \frac{\widehat{z}^2}{8} \right) +
B_1 \, \widehat{z}^{\,-\frac{1}{2}} \left(1+\frac{\widehat{z}^2
\ln(1/\widehat{z})}{2}\right)\right] & \text{for} \, \; \alpha=1\,.
\end{cases}
\end{equation}
In fact, the $\widehat{z}^2$ correction to the $\widehat{z}^{\,\frac{1}{2}
+\alpha}$ solution will not play any role and we drop it in the following. 
With this simplification, Eq.~(\ref{smallzcorrect}) together with the 
UV boundary conditions implies
\bea 
\label{balphatuned}
B_\alpha &=& 
4\alpha(1-\alpha)\widehat{z}_{\uv}^{\,2\alpha-2}\,\,\,,\label{bat}\\ 
\nonumber\\
B_1 &=& \frac{4}{2\ln(1/\widehat{z}_{\uv})+1}\,\,.\label{b1}
\eea

This result can be intuitively understood as follows: It is clear that, 
for generic $\lambda$, the boundary condition Eq.~\eqref{fullBCz} can 
only be fulfilled if both solutions to the wave equation, 
$\widehat{z}^{\,\frac{1}{2} + \alpha}+\cdots$ and $B_{\alpha} \, 
\widehat{z}^{\,\frac{1}{2} - \alpha}+\cdots$, are of comparable size at the 
UV brane. This immediately gives $\smash{B_\alpha \sim \widehat{z}_\uv^{\,2 
\alpha}}$ as in Eq.~\eqref{balphadetuned}. Now, for tuned $\lambda_0=2 - 
\alpha$, the function $B_\alpha\,\widehat{z}^{\,\frac{1}{2} - \alpha}$ 
fulfills the boundary condition already by itself, as can be easily checked. 
Therefore, the boundary condition now forces the leading correction to that 
function, $B_{\alpha} \, \widehat{z}^{\,\frac{5}{2} - \alpha}$, to have the 
same size as the other solution, $\widehat{z}^{\,\frac{1}{2} + \alpha}$, at 
the UV brane. This gives $\smash{B_\alpha \sim \widehat{z}_\uv^{\,2\alpha-2}}$
in agreement with Eq.~\eqref{balphatuned}.

We now observe that, as long as $\alpha<1$, the 
$\widehat{z}^{\,\frac{1}{2}-\alpha}$ solution is dominant everywhere 
between $\widehat{z}=\widehat{z}_{\uv}$ and $\widehat{z}={\cal O}(1)$.
This justifies our previously derived rate of Eq.~(\ref{0a1}) and 
shows that the presently discussed $\widehat{z}^2$ corrections are
not important for $0\leq\alpha<1$.

Next, we focus on the regime $\alpha>1$. In this case, we see that the 
$\widehat{z}^{\,\frac{1}{2}-\alpha}$ solution dominates for 
$\widehat{z}_{\uv}<\widehat{z}<\widehat{z}_{\uv}^{\,1-1/\alpha}$.
By contrast, the $\widehat{z}^{\,\frac{1}{2}+\alpha}$ solution is larger 
for $\widehat{z}_{\uv}^{\,1-1/\alpha}<\widehat{z}<{\cal O}(1)$. Although 
Fig.~\ref{sketch} does not describe this regime correctly, one can check that 
the normalization is nevertheless dominated by the IR region. It is then easy 
to see that 
\be
\label{psi1a2}
\psi_n(\widehat{z}_{\uv})\sim z_{\uv}^{\,\alpha-\frac{3}{2}}
m_{\ir}^{\frac{1}{2}}\qquad\text{for} \quad \alpha>1 \quad \mbox{and} \quad 
\lambda=\lambda_0\,,
\ee
giving the decay rate
\be
\label{a12}
\Gamma_n\sim\left(\frac{m_n}{k}\right)^{2\alpha}m_{\ir}\qquad
\text{for} \quad \alpha>1 \quad \mbox{and} \quad \lambda=\lambda_0\,.
\ee
It can be checked that, in spite of the extra logarithms appearing at
intermediate steps, the special case $\alpha=1$ is correctly reproduced
by simply taking the appropriate limit of either Eq.~(\ref{0a1}) or 
Eq.~(\ref{a12}). 

The 5d graviton with indices $\mu,\nu$ has the same equation of motion as a 
massless 5d scalar without boundary mass (i.e.~for $\lambda = \lambda_0 = 0$). 
We assume $M_5 \sim k$ in the following. The coupling of the
5d graviton to the energy-momentum tensor on the UV brane is then 
suppressed by a factor $k^{-3/2}$ as in Eq.~\eqref{assumedcoupling}. 
Hence Eq.~(\ref{a12}) with $\alpha=2$ also applies to the graviton and we 
have 
\begin{equation}
\label{masslessdecayrate2}
\Gamma_n \, \sim \,
\left(\frac{m_n}{k}\right)^{4} m_{\ir}\qquad\text{for graviton KK modes} \,,
\end{equation}
in agreement with the literature on graviton tunneling between two throats 
(see e.g. \cite{Dimopoulos:2001ui,Langfelder:2006vd,
Harling:2007jy}). These results are relevant since the throat to which the 
KK modes decay is dual to a UV-brane-localized gauge theory 
\cite{Harling:2007jy}. Of course, Eq.~(\ref{masslessdecayrate2}) corresponds 
to the case where this gauge theory has ${\cal O}$(1) degrees of freedom. 
We finally note that the coupling strength of graviton KK modes to the 
UV brane has been given, e.g., in App. A of \cite{Chung:2000rg}. 
For $M_5 \sim k$, it reads $\smash{g_n \sim \sqrt{m_n m_\ir}/k^2}$. The 
decay rate then follows from Eq.~\eqref{decayrate} and also reproduces
Eq.~\eqref{masslessdecayrate2}.

\section{Situations dominated by the tachyon}
\label{tachyondominated}
A detailed discussion of the application of our 5d analysis to
the Klebanov-Strassler throat is the subject of 
Sect.~\ref{ApplicationstoThroats}. However, to explain the relevance of 
tachyon decay rates we have to jump somewhat ahead and mention certain 
facts concerning warped flux compactifications already in this section: 
In flux compactifications, scalars obtain a large mass in the 
unwarped part of the compact space.\footnote{We 
assume that the K\"ahler moduli are stabilized as well (which requires 
effects other than flux).
}
In our 5d model, this corresponds to a large and generically detuned mass 
term on the UV brane. It is then evident from Eq.~\eqref{drs} that a 
maximally tachyonic scalar ($\alpha=0$) has a considerably larger decay 
rate than a massless scalar ($\alpha=2$). This can be easily understood from 
the quantum mechanical analog: The suppression effect of the UV mass term is 
present for both fields, but only the massless scalar has to surmount an 
additional potential barrier. Thus, assuming that a certain KK mode decays
only via scalars and that all relevant scalars have a large UV-brane mass
term, the tachyon governs the decay. Clearly, the assumption that decays 
proceed only through scalars is non-trivial: The graviton decay rate, 
Eq.~\eqref{masslessdecayrate2} is larger than the tachyon rate, 
Eq.~\eqref{drs} (albeit only by a factor $\smash{(\ln(k/m_n))^{-2} \sim 
(k \ell)^{-2}}$). However, this is only relevant for spin-two KK modes 
which can mix with the graviton. 

Next, we observe that there are specific situations where, in contrast 
to what was said above, the tachyon decay rate is larger than that of the 
graviton. This is possible because the tachyon wavefunction rises while the 
graviton wavefunction falls when moving away from the UV brane. Thus, if 
there is a probe brane in the throat which is localized somewhere between the 
UV and IR brane, the decay to this probe brane can be dominated by the tachyon
(even if a corresponding graviton decay is allowed). We use the simple ansatz 
Eq.~\eqref{assumedcoupling} for the coupling to gauge fields on the probe 
brane at $\widehat{z}_\delta$ (where $\widehat{z}_{\uv}<\widehat{z}_\delta<
\widehat{z}_{\ir}$).\footnote{As 
a string-theoretic realization of this situation, consider a D7 brane
wrapping a 3-cycle of the $T^{1,1}$ in a Klebanov-Strassler throat and 
extending from the compact Calabi-Yau to a certain lowest point in the throat. 
For a discussion of such embeddings and applications see e.g. \cite{
Kuperstein:2004hy,Gherghetta:2006yq,Levi:2005hh,Chen:2008jj}. Even though, 
in 5d language, the D7-brane gauge theory lives everywhere between 
$\widehat{z}_{\uv}$ and $\widehat{z}_\delta$, we model this situation 
by a gauge theory localized at $\widehat{z}_\delta$. This is a reasonable 
approximation since the coupling to throat fields is dominated by 
interactions at the largest relevant values of $\widehat{z}$.}

We now demonstrate the enhancement of the decay rates quantitatively. Let 
the probe brane be localized at $y = \delta$ for some $\delta \ll \ell$. 
For the light modes, $\widehat{z}_\delta = \widehat{z}_{\uv} \, e^{k \delta} 
\sim n  e^{- k(\ell-\delta)}$ is small and we can use the approximate 
wavefunctions of Eq.~\eqref{smallzsol}. We see that, as $\widehat{z}_\delta$ 
increases, the tachyon wave function ($\alpha=0$) grows 
like $\widehat{z}^{\,\frac{1}{2}}$ (we neglect the additional logarithmic 
dependence). By contrast, the KK graviton wave function ($\alpha=2$) falls
like $\widehat{z}^{\,\frac{1}{2}-\alpha}$. This is because, as explained 
before Eq.~(\ref{psi1a2}), the falling solution dominates near the UV brane. 
Thus, the ratio of the two wave functions at $\widehat{z}_\delta$ is enhanced, 
relative to its UV brane value, by a factor $(\widehat{z}_\delta/
\widehat{z}_{\uv})^2$. Given that the tachyon decay rate to the UV brane 
is suppressed relative to the corresponding graviton rate by the logarithmic 
factor mentioned above, we conclude that 
\be
\frac{\Gamma_{\rm tachyon}}{\Gamma_{\rm graviton}}\sim
\frac{(\widehat{z}_\delta/\widehat{z}_{\uv})^4}{\ln^2(\widehat{z}_{\ir}/
\widehat{z}_{\uv})}
\ee
for decays to the probe brane. Thus, the tachyon starts to dominate the 
decay when the probe-brane warp factor $(\widehat{z}_\delta/
\widehat{z}_{\uv})^{-1}$ becomes smaller than $1/\sqrt{\ln(\widehat{z}_{\ir}/
\widehat{z}_{\uv})}$. This is obviously a very weak requirement.

Finally, we mention a third situation in which the tachyon dominates the
decay rates to the UV sector. We can not argue from the 5d perspective 
how natural or unnatural it is to have a tuned UV mass term, $\lambda=
\lambda_0$. (We will return to this question in the string-theoretic 
context in Sect.~\ref{ApplicationstoThroats}.) However, a tuned value of 
$\lambda_0$ is clearly a legitimate possibility. Combining Eqs.~(\ref{0a1}) 
and (\ref{a12}), the corresponding decay rates can be written as
\be
\Gamma_n\sim \left(\frac{m_n}{k}\right)^{2+2|\alpha-1|}m_{\ir} \qquad 
\text{for}\quad \lambda=\lambda_0\,.
\ee
For $0<\alpha<2$, this obviously dominates the decay rates corresponding to 
massless scalars and the graviton. The maximal enhancement is realized for
$\alpha=1$, i.e., for a tachyon the negative mass-squared of which is half 
that of the Breitenlohner-Freedman bound.

\section{CFT Interpretation}
\label{cftpicture}

We now present physical arguments for the decay rates of 
Sect.~\ref{adsdecays} using the `CFT-dual' description of a 
Randall-Sundrum model as a strongly coupled 4d field 
theory. In other words, we are going to appeal to a simplified,
purely field-theoretic version of the AdS/CFT correspondence \cite{
Maldacena:1997re}. In this description, a scalar bulk field with mass 
squared $M^2$ corresponds to a CFT operator $\mathcal{O}_\Delta$ with 
conformal dimension $\Delta = 2 + \sqrt{4 + M^2/k^2} = 2 + \alpha$.\footnote{As 
we will discuss below, for 
$\alpha <1$, also the relation $\Delta = 2 -\alpha$ can be realized 
\cite{Klebanov:1999tb}.
} 
Without a UV brane, the generating functional of CFT correlators can be 
obtained from the gravity description via the relation \cite{Gubser:1998bc,
Witten:1998zw}
\begin{equation}
\label{adscftquant1}
 \int \! \! \! \; \mathcal{D} \phi_{\cft} \, e^{- S_{\cft}[\phi_{\cft}] - \int
d^{4}x \, \phi_0
\mathcal{O}_\Delta} =
\int_{\phi_0} \! \! \! \mathcal{D}\Phi
\, e^{- S_{5d}[\Phi]} \; .
\end{equation}
Here $\phi_{0}(x)$ specifies the behaviour of the bulk field $\Phi(x,z)$ near the boundary of 
AdS$_5$ (taking $z \to 0$) to be $\Phi(x,z) \simeq \phi_{0}(x) \, z^{4 - \Delta} \, k^{3/2}$. It thus acts as a 
source for the corresponding operator in the generating functional.\footnote{More precisely, $\Phi(x,z) \simeq 
\phi_{0}(x) \, z^{4 - \Delta} \, k^{3/2} + A(x) \, z^{\Delta} \, k^{3/2}$, where $A(x)$ can be interpreted as the 
expectation value of the dual operator $\mathcal{O}_\Delta$ and $\phi_{0}(x)$ as the source. The 
factors $k^{3/2}$ have been introduced for dimensional reasons.} The value 
of $\phi_0$ is kept fixed in the functional integral on the right-hand side
and $\phi_{\cft}$ denotes the dynamical CFT fields.

In the Randall-Sundrum model, there is a brane at $\smash{z=
k^{-1}}$, corresponding to a UV cutoff $\Lambda_{\uv} = k $ in the CFT. The source 
$\smash{\phi_0(x)= k^{5/2-\Delta} \Phi(x,k^{-1})}$ now becomes a physical degree of freedom and has to 
be included in the functional integral \cite{Gherghetta:2006ha}:
\begin{equation}
\label{adscftquant}
\int \! \mathcal{D}\phi_0 \, e^{- S_{\uv}[\phi_0]} \int_{\Lambda_{\uv}} 
\! \! \! \mathcal{D}\phi_{\cft} \, e^{- S_{\cft}[\phi_{\cft}] - \int d^{4}x 
\, \phi_0 \mathcal{O}_\Delta} = \int \! \mathcal{D}\phi_0 \, 
e^{- S_{\uv}[\phi_0]}\int_{\phi_0} \! \! \! \mathcal{D}\Phi
\, e^{- S_{5d}[\Phi]} \, .
\end{equation}
We may think of the l.h. side of Eq.~(\ref{adscftquant}) as being {\it 
defined} by the r.h. side. In our case, the UV-brane action of the bulk 
scalar field, $S_{\uv}[\phi_0] $, consists of the UV-brane mass term and the 
interaction term with the brane-localized gauge theory. (We have suppressed 
the gauge fields in the functional integral in Eq.~(\ref{adscftquant}) for 
notational simplicity.) Thus, at the UV scale $\Lambda_{\uv}$, the complete 
Lagrangian on the CFT side of the duality reads
\begin{equation}
\label{cftlag}
\mathcal{L} \, = \, \mathcal{L}_{\cft}[\phi_{\cft}] + M_{\phi_0}^2 \, 
\phi_{0}^2 + \frac{1}{\Lambda_{\uv}^{\Delta - 3}} \, \phi_0 \,
\mathcal{O}_\Delta + \frac{1}{\Lambda_{\uv}} \, \phi_0 \,\mbox{tr}
\left(F^{\mu \nu} F_{\mu \nu}\right)\,,
\end{equation}
where we have redefined $\phi_0$ by an appropriate power of 
$\Lambda_{\uv}$ to give it mass dimension one. 

When running the action corresponding to Eq.~(\ref{cftlag}) down to scales 
below $\Lambda_{\uv}$, the CFT induces a kinetic term $\sim 
(\partial\phi_0)^2$ for the scalar $\phi_0$, which now manifestly becomes a 
propagating field \cite{PerezVictoria:2001pa}. Similarly, the tachyonic 
instability which arises for $M^2<0$ in the presence of a UV brane can be 
understood as an effect of CFT-induced corrections. In this language, 
loop effects of the CFT drive a scalar $\phi_0$ with $M_{\phi_0}^2=0$ at 
the UV scale to a negative mass squared at the IR scale if $\Delta < 4$ 
\cite{Delgado:2003tx}. A sufficiently large mass term at the UV scale 
prevents the CFT from making the scalar tachyonic. In particular, the tuning 
$\lambda = \lambda_0$ of the UV mass described earlier can be interpreted in 
the CFT language as a tuning of $M_{\phi_0}^2$ which ensures that $\phi_0$ becomes 
massless at the IR scale.

We first focus on the case of a generic, large UV mass term, 
$\lambda = \lambda_{0} + \mathcal{O}(1)$. In this case, $\phi_0$ has a mass 
of the order $\Lambda_{\uv}$ even after running down to the IR scale 
$m_{\ir}$. Since $\phi_0$ remains heavy all the way from $\Lambda_{\uv}$ 
down to $m_{\ir}$, its effect on the CFT dynamics and, in particular, on 
the dimension of $\mathcal{O}_\Delta$ is negligible. Thus, the IR scale
Lagrangian is still that of Eq.~\eqref{cftlag}, modified only by 
a kinetic term and a mass correction for $\phi_0$. We can now integrate 
out $\phi_0$ entirely, finding the coupling 
\begin{equation}
\label{coupling}
\sim\frac{1}{\Lambda_{\uv}^{\Delta - 2}} \, \frac{1}{M_{\phi_0}^{2}} \,
\mathcal{O}_\Delta \,\mbox{tr}\left(F^{\mu \nu} F_{\mu \nu}\right)
\end{equation}
between the brane-localized gauge theory and the operator 
$\mathcal{O}_\Delta$. At low scales, the conformal symmetry of 
${\cal L}_{\rm CFT}$ is broken by the IR brane. In the case of the 
Klebanov-Strassler throat, this breaking of (approximate) conformal 
invariance is a dynamical effect within the strongly-coupled gauge theory. 
The low-lying glueball states of this gauge theory correspond to (linear 
combinations \cite{Batell:2007jv} of) the KK modes of the bulk fields. 
Assuming that these glueballs have non-vanishing 
overlap with ${\cal O}_\Delta$, they decay to the brane-localized gauge 
theory via the operator in Eq.~\eqref{coupling}. Since we assumed that 
$M^2_{\phi_0} \sim \Lambda^2_{\uv}$ even at the IR scale, the decay rate 
\begin{equation}
\label{cftalphalarger1massivephi0}
\Gamma \sim \displaystyle \left(\frac{m_{\ir}}{\Lambda_{\uv}}\right)^{2
\Delta} m_{\ir}
\end{equation}
follows by dimensional analysis. Recalling that $\Lambda_{\uv}=k$ and 
$\Delta = 2 + \alpha$, we see immediately that this agrees with the 
first line of Eq.~(\ref{drs}).

We now turn our attention to the tuned case, $\lambda = \lambda_0$,
making use of the holographic picture developed in 
\cite{Gherghetta:2006ha,Delgado:2003tx}. As before, we run the Lagrangian 
of Eq.~(\ref{cftlag}) down to the scale $m_{\ir}$. In contrast to our 
previous discussion, we can not appeal to the large mass of $\phi_0$ to 
ignore its influence on the CFT. Indeed, as explained earlier, 
$M_{\phi_0}^2$ approaches zero as the energy scale approaches $m_{\ir}$. 
However, provided that $\alpha>1$ ($\Delta>3)$, we can now argue that the 
influence of $\phi_0$ is negligible because of its small mixing with CFT 
states. Indeed, for $\Delta>3$, the operator $\phi_0{\cal O}_\Delta$ 
is irrelevant. Thus, we assume that the IR scale effective action contains 
light glueballs and a light field $\phi_0$, with a mixing term suppressed 
by $1/\Lambda_{\uv}^{\Delta-3}$. 

\begin{figure}[t]
\begin{center}
\includegraphics[scale=0.47]{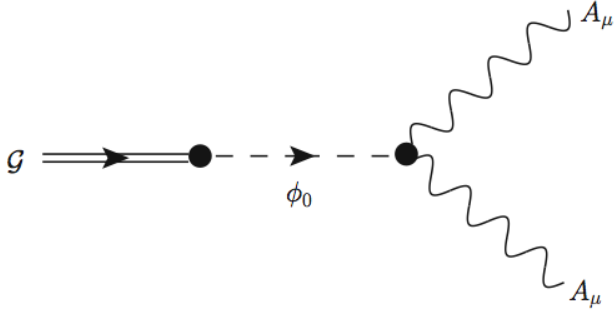}
\caption{Decay of a glueball $\mathcal{G}$ 
via the light field $\phi_0$ into two gauge fields $A_\mu$.}
\label{decayviamasslessphi}
\end{center}
\end{figure}

The decay of glueballs to fields of the UV-brane gauge theory now proceeds 
via the massless or light `source field' $\phi_0$, cf. 
Fig.~\ref{decayviamasslessphi}. The rate follows by dimensional analysis:
\begin{equation}
\label{cftalphalarge1masslessphi0}
\Gamma \sim \left(\frac{m_{\ir}}{\Lambda_{\uv}}\right)^{2 \Delta - 4} 
m_{\ir} \, .
\end{equation}
This result can also be immediately obtained from 
Eq.~(\ref{cftalphalarger1massivephi0}): That equation secretly contains
a suppression factor $1/(\Lambda_{\uv}^2)^2$ coming from the massive 
$\phi_0$ propagator. Replacing this mass by $m_{\ir}$, i.e. multiplying 
Eq.~(\ref{cftalphalarger1massivephi0}) by $(\Lambda_{\uv}/m_{\ir})^4$, we
obtain Eq.~(\ref{cftalphalarge1masslessphi0}). This rate agrees with 
the result of the gravity calculation for $\alpha > 1$ given in Eq.~(\ref{a12}).

For $\alpha<1$ ($\Delta<3$) the above line of argument breaks down since the mixing
between $\phi_0$ and the CFT states can not any more be considered a
small perturbation. This can also be understood from our gravity-side
discussion in Sect.~\ref{adsdecays}. Indeed, we are dealing with a
situation where the UV-brane mass term is tuned to allow for an exact
zero mode. The bulk profile of this zero-mode is $\psi_0 \sim z^{\,\frac{1}{2}-\alpha}$.
Thus, its normalization is UV-localized for $\alpha>1$ and IR-localized
for $\alpha<1$ (cf.~Eq.~\eqref{normalization}). 
The corresponding 4d field can then be viewed as a
UV-brane- or IR-brane-localized field respectively. In the case
$\alpha>1$, this supports our previous statement that $\phi_0$ does not
affect the CFT dynamics significantly. By contrast, in the $\alpha<1$
case, the zero mode is tangled up in a non-trivial way in the
non-perturbative CFT dynamics \cite{Gherghetta:2006ha,Batell:2007jv}.

In order to understand the case $\alpha<1$ with tuned $\lambda=\lambda_0$ from the CFT
perspective, we will first describe the analysis of this section from an 
equivalent but technically slightly different point of view.
When running the Lagrangian of Eq.~(\ref{cftlag}) down
to smaller scales, we are consecutively integrating out $\phi_0$ modes 
with lower and lower 4-momenta $k$. This induces corrections which are
schematically of the form
\be
\label{o2}
\sim \frac{1}{\Lambda_{\uv}^{2(\Delta - 3)}} \frac{{\cal O}_\Delta(k)^2}{M_{\phi_0}^2(k)+c(k)k^2}\,.
\ee
Here the function $c$ is zero at the high scale, $c(\Lambda_{\uv})=0$, 
and grows at lower scales to the extent that a kinetic term for $\phi_0$ is 
induced by CFT loops. If $M_{\phi_0}^2(k)$ becomes small and the induced operator 
$\smash{\sim {\cal O}_\Delta{}^2}$ is not suppressed by $\Lambda_{\uv}$ (i.e.~for $\Delta<3$), it
can be large enough to 
modify the CFT dynamics significantly and to invalidate the derived decay 
rate. Moreover, the effect of this operator must be the origin of the
tachyonic instability which is present in the CFT for $\lambda<\lambda_0$.

Alternatively,
we may immediately integrate out $\phi_0$, before considering any RG running or 
loop effects. This induces
a coupling of the type given in Eq.~(\ref{coupling}) 
(but with $M_{\phi_0}^2=M_{\phi_0}^2(\Lambda_{\uv})$). It also induces the term
\be
\label{o3}
\sim \frac{{\cal O}_\Delta^2}{(\Lambda_{\uv})^{2(\Delta-3)}M_{\phi_0}^2}
\ee
which, in contrast to the non-local corrections given schematically in 
Eq.~(\ref{o2}), is a local operator. In other words, the presence of 
a UV brane with a brane-localized mass term can be viewed, on the 
CFT side, as a correction by a `double trace operator' as discussed in
\cite{Witten:2001ua}.

It is now immediately clear that, as $M_{\phi_0}^2$ is large in the detuned case,
this ${\cal O}_\Delta^2$ correction is too small to affect the CFT
dynamics at low energy scales (and in particular the dimension of ${\cal 
O}_\Delta$) significantly. Hence, the coupling ${\cal O}_\Delta F^2$ can be 
directly used to estimate the decay rate of any light glueball with non-vanishing
overlap with ${\cal O}_\Delta$. This reproduces Eq.~(\ref{cftalphalarger1massivephi0}).

It is also clear that, from this perspective, the value $\Delta=2$ (corresponding 
to ${\alpha=0}$) deserves special attention. Indeed, in this case the ${\cal O}_\Delta^2$
operator is marginal rather than irrelevant. As discussed in Sect.~4 of 
\cite{Witten:2001ua}, this causes multiplicative renormalization of the operator 
${\cal O}_\Delta$ with logarithmic running:
\be
{\cal 
O}_\Delta(m_\ir) \, \sim \, \ln(\Lambda_\uv/m_\ir) \, {\cal O}_\Delta(\Lambda_\uv) \, .
\ee
Upon rescaling the operator in Eq.~\eqref{coupling}, we obtain an additional factor of 
$\smash{\ln(\Lambda_\uv/m_\ir)^{-2}}$ in the decay rate Eq.~\eqref{cftalphalarger1massivephi0}, 
in agreement with the factor $\smash{\ln(k/m_\ir)^{-2}}$ in the second line of Eq.~(\ref{drs}).

We now consider the case $\alpha<1$ with tuned
boundary mass, $\lambda=\lambda_0$, corresponding to the largest
consistent coefficient of the ${\cal O}_\Delta^2$ correction. From the
discussion in Sect.~\ref{adsdecays} (see in particular 
Eqs.~(\ref{smallzsol}) and (\ref{balphatuned})) we see that, in this case, 
the UV behaviour of the AdS scalar changes from $\smash{\Phi\propto z^{2+\alpha}}$ to 
$\smash{\Phi\propto z^{2-\alpha}}$. According to the general discussion of 
\cite{Klebanov:1999tb}, the dual description of an AdS field theory 
with such boundary conditions and with $\alpha<1$ is provided by a CFT 
with an operator ${\cal O}_{\Delta'}$ with dimension $\Delta'=2-\alpha$ 
(instead of $\Delta=2+\alpha$). When we integrate $\phi_0$ out at the UV scale, we obtain
the coupling
\be
\label{o4}
\sim \frac{1}{\Lambda_{\uv}^{\Delta'}} \, {\cal O}_{\Delta'} \,\mbox{tr}
\left(F^{\mu \nu} F_{\mu \nu}\right)
\ee
as well as a term of the type given in Eq.~\eqref{o3}. One may in fact say that it is this latter 
correction with an appropriately tuned coefficient which forces the dimension of ${\cal O}$ to change to 
$\Delta'=4-\Delta$ in the infrared 
\cite{Kaplan:2009kr} (see also \cite{Witten:2001ua,Moroz:2009kv}). 
From Eq.~\eqref{o4}, we can estimate the decay rate of glueballs with non-vanishing 
overlap with ${\cal O}_{\Delta'}$ as
\begin{equation}
\Gamma \sim \left(\frac{m_{\ir}}{\Lambda_{\uv}}\right)^{2 \Delta'}
m_{\ir} \,,\label{drp}
\end{equation}
in agreement with Eq.~(\ref{0a1}).

While a more systematic and quantitative study of the CFT description of
the decay rates under consideration may be worthwhile, we 
believe that we have now supplied enough additional physical intuition for 
the purposes of the present paper. In particular, we have fully confirmed 
the results of our gravity-side calculation.

\section{Application to throats in flux compactifications}
\label{ApplicationstoThroats}

We now discuss the applicability of our results to throat geometries 
in type IIB string theory. For definiteness, we focus on the 
Klebanov-Strassler (KS) throat \cite{Klebanov:1998hh,Klebanov:1999rd,
Klebanov:2000nc,Klebanov:2000hb} in the following. We expect, however, that 
our results can similarly be applied to other types of throats.

The curvature scale $R=k^{-1}$ of the KS throat varies logarithmically 
along the radial direction and the geometry smoothly terminates in the IR. We 
neglect this logarithmic variation in the following. Sufficiently far 
away from the IR tip, the KS throat can then be approximated by the space
AdS$_5 \times T^{1,1}$. In the UV, the throat is smoothly glued into a 
compact manifold~\cite{Giddings:2001yu}. We focus on compactifications 
in which the size $L$ of this manifold is not hierarchically larger than the 
AdS scale $R$ of the throat.\footnote{The 
throat can only be glued smoothly into the compact manifold if $L \gtrsim R$. 
The distribution of flux vacua strongly favors vacua where the volume 
$L^6$ is small in string units, cf.~\cite{Denef:2008wq} and references 
therein. Thus, the generic compactification has no large hierarchy between 
$R$ and $L$.
} 
We can then neglect the `thickness' of the compact manifold and approximate 
it by the UV brane of an RS model. Similarly, we model the IR end of the 
throat by an IR brane (see 
Ref.~\cite{Brummer:2005sh} for more details on the KS throat as a RS model).

The KK reduction of type IIB supergravity on $T^{1,1}$ in an AdS$_5 \times
T^{1,1}$ background has been carried out in \cite{Ceresole:1999zs}.
The resulting spectrum of KK modes contains scalars with various 
tachyonic masses, some of which saturate the Breitenlohner-Freedman (BF) 
bound. Note that the presence of scalars saturating this bound is expected on 
general grounds: The theory in AdS$_5$, which results from the KK reduction, 
is supersymmetric. Furthermore, the isometries of $T^{1,1}$ ensure the 
presence of massless vector multiplets in the spectrum. Such multiplets 
contain a scalar with maximally tachyonic mass \cite{Kim:1985ez,
Shuster:1999zf,Gherghetta:2000qt}.

We will now discuss tachyons from such multiplets in more detail. The 
KK reduction of type IIB supergravity on AdS$_5\times T^{1,1}$ contains 
eight massless 5d vectors. One of these vectors comes from the 
compactification of the 4-form 
potential on the 3-cycle in $T^{1,1}\sim S^3\times S^2$. The corresponding 
abelian symmetry of the solution is called $U(1)_B$. The seven remaining 
vectors are associated with the $SO(4)\times U(1)_R$ isometry of $T^{1,1}$.
Since the $U(1)_R$ vector is part of the massless graviton multiplet 
\cite{Ceresole:1999zs} which contains no scalars (and in particular no 
tachyon), we end up with seven BF tachyons in total. In the classification 
of \cite{Ceresole:1999zs}, these scalars belong to shortened versions of 
vector multiplet I.

In a KS throat, the symmetries are reduced with respect to AdS$_5\times T^{1,1}$.
In the region between the IR and UV end, which is the most symmetric part of 
the throat, 
the symmetry is reduced to $SO(4) \times U(1)_B \times \mathbb{Z}_{2M}\subset 
SO(4) \times U(1)_B \times U(1)_R$. The $U(1)_R$ is broken to $\mathbb{Z}_{2M}$ 
by the $M$ units of 3-form-flux on 
the $S^3$, which are also responsible for the logarithmic variation of the 
AdS scale \cite{Klebanov:2002gr}. Note, however, that this breaking does not 
affect our previous counting of BF tachyons since, as mentioned before, the 
$U(1)_R$ vector has no scalar partner.

The symmetry is further reduced at the IR and UV end of the KS throat: In the IR, 
the $SO(4)$ stays intact, 
but the $U(1)_B$ is completely broken, and the $\mathbb{Z}_{2M}$ is broken to 
$\mathbb{Z}_2$ \cite{Klebanov:2000hb,Aharony:2000pp,Gubser:2004qj,Gubser:2004tf}. 
In the UV, on the other hand, the $SO(4)\times U(1)_R$ isometry of $T^{1,1}$ 
is broken since a compact Calabi-Yau has no continuous isometries. Concerning 
the $U(1)_R$ factor, we 
are anyway only interested in its $\mathbb{Z}_2$ subgroup which survives in 
the IR region. This discrete $\mathbb{Z}_2$ symmetry (or a larger discrete 
subgroup of the $T^{1,1}$ isometry) may or may not be respected by the 
compact Calabi-Yau. The latter is certainly the generic case.
In addition, the KS throat has another $\mathbb{Z}_2$ symmetry (called 
$\mathcal{I}$-symmetry in \cite{Gubser:2004tf}) which interchanges the 
two 2-sphere in the $T^{1,1}$ ($T^{1,1}$ is an $S^1$ bundle over $S^2 
\times S^2$) and reverses the sign of the 
2-form potentials of type IIB supergravity\cite{Klebanov:1998hh,Morrison:1998cs}. 
This symmetry is unbroken
in the IR and may or may not be broken in the UV.\footnote{As long as the 
throat is infinite, we can move along a one-parameter family of 
solutions, the so-called baryonic branch of the dual gauge theory, and the KS 
solution is a special point which respects the $\mathcal{I}$-symmetry 
\cite{Gubser:2004qj,Gubser:2004tf,Papadopoulos:2000gj,Butti:2004pk}. However, the 
metric deformation which corresponds to moving along the baryonic branch does not 
respect the conformal-Calabi-Yau condition \cite{Butti:2004pk}. Hence, it is not 
clear how to glue such a deformed throat into a compact UV space in the framework 
of warped flux compactifications \cite{Giddings:2001yu}. For this reason, we do not 
consider such deformations. It is, of course, 
nevertheless possible that the $\mathcal{I}$-symmetry is broken by the 
Calabi-Yau (instead of the throat), corresponding to a $\mathbb{Z}_2$-breaking on the UV 
brane.}
Furthermore, the $U(1)_B$ can be broken or remain unbroken in the UV. This is 
not essential 
for us since the corresponding BF scalar, being in the adjoint, is uncharged.
Thus, the couplings of this scalar to various UV localized fields, which 
are our main interest, are not forbidden by this symmetry. We finally note 
that the symmetry breaking in the UV is mediated to the IR by irrelevant 
operators and is thus suppressed by powers of the warp factor~\cite{
DeWolfe:2004qx,Aharony:2005ez,Berndsen:2008my,Dufaux:2008br}.\footnote{As 
discussed in \cite{Baumann:2008kq}, relevant perturbations (which grow towards
the IR) are also possible 
as long as they start with a sufficiently small amplitude in the UV.}

We now discuss specifically the effects of the tachyon in the $U(1)_B$ vector 
multiplet (also known as Betti multiplet). This tachyon is an $SO(4)$ singlet, 
but odd under the $\mathcal{I}$-symmetry 
$\mathbb{Z}_2$~\cite{Gubser:2004qj}.\footnote{We 
note that, in the infinite throat limit, this 5d scalar has a massless 4d 
mode \cite{Gubser:2004qj,Benna:2007mb}. When the throat is glued into a 
compact manifold, this mode obtains a mass which is parametrically large 
compared to the IR scale $m_{\ir}$ if the volume of the compact manifold is 
small. Thus, this mode is not important in our context.
}
In the IR region of the KS throat, where the approximation as AdS$_5 
\times T^{1,1}$ becomes unreliable, this tachyon mixes with a scalar of 
mass $M^2=5 k^2$ \cite{Benna:2007mb,Dymarsky:2008wd}. In the UV,
on the other hand, the two scalar fluctuations decouple, as we will now 
demonstrate. The equations of 
motion of this system are given in Eqs.~(47) and (48) in \cite{Benna:2007mb}:
\begin{gather}
\tilde{z}'' - \frac{2}{\sinh^2\tau} \tilde{z} +\tilde{m}^2 
\frac{I(\tau)}{K^2(\tau)}\tilde{z} \, 
= \, \tilde{m}^2 \frac{9}{4 \cdot 2^{2/3}} K(\tau) \tilde{w} \\
\tilde{w}'' - \frac{\cosh^2\tau+1}{\sinh^2\tau} \tilde{w} +\tilde{m}^2 
\frac{I(\tau)}{K^2(\tau)}\tilde{w} \, 
= \, \frac{16}{9} K(\tau) \tilde{z} \, .
\end{gather}
Here, primes denote derivatives with respect to the radial coordinate $\tau$ 
of the KS throat (using the metric convention of \cite{Klebanov:2000hb}) and 
$\tilde{m}$ is related to the 4d mass of the state.
The functions $I(\tau)$ and $K(\tau)$ are e.g. given in Appendix B
in \cite{Benna:2007mb}. In the UV, $\tau \gg 1$, these functions can be 
approximated by $\smash{I(\tau) \sim \tau \, e^{-4 \tau/3}}$ and 
$\smash{K(\tau) \sim e^{-\tau/3}}$. The equations of motion then 
indeed decouple and simplify to
\begin{gather}
\tilde{z}'' \, = \, 0 \\
\tilde{w}'' \, - \, \tilde{w} \, = \, 0 \,.
\end{gather}
This is solved by $\tilde{z} =  \tau + const.$ and 
$\smash{\tilde{w} = e^\tau + const. \, e^{-\tau}}$ (neglecting the overall 
factors). For $\tau \gg 1$, the coordinate $\tau$ is related to our 
coordinate $z$ by $\smash{z \propto e^{-\tau/3}}$ and the wavefunctions 
read $\tilde{z} =  \ln z + const.$ and 
$\smash{\tilde{w} = z^{-3} + const. \, z^3}$. Up to an overall factor of $z^2$ 
(which is related to a field redefinition\footnote{In particular, the field $\tilde{z}$ 
is related to the fluctuation $\psi = \delta g_{13} = \delta g_{24}$ of the  
5-dimensional compact manifold in the throat by the field redefinition 
$\smash{\tilde{z} = z^{-2}\psi}$. In the UV of the 
throat, where the approximation as AdS$_5 \times T^{1,1}$ is applicable,
the field $\psi$ is a 5d scalar with a standard kinetic term (coming from the 
10d Einstein-Hilbert term).}) these are indeed the 
wavefunctions of a tachyon which saturates the BF bound and a scalar with mass
$M^2= 5 k^2$. In particular, we see that the wavefunctions from the KK 
decomposition on AdS$_5 \times$T$^{1,1}$ are a good approximation in the 
UV of the KS throat.

The mass spectrum of 
4d KK modes from this system of two scalars was determined in \cite{Benna:2007mb} and
found to contain the lightest state which is known so far for the KS throat 
(see e.g.~\cite{Krasnitz:2000ir,Caceres:2000qe,Amador:2004pz,Noguchi:2005ws,
Berg:2006xy,Dymarsky:2007zs} for other parts of the mass spectrum). Since 
heavier KK modes decay very quickly to lighter states via various processes 
(which we will discuss in Sect.~\ref{cosmology}), this lightest KK mode 
generically contains an $\mathcal{O}(1)$-fraction of the energy density of a 
heated KS throat.\footnote{More 
precisely, this state is part of a massive vector multiplet of 4d
$\mathcal{N}=1$ supersymmetry \cite{Dymarsky:2008wd}. A similar fraction 
of the energy density is therefore stored in the superpartners of this 
scalar.
} 
Due to the effect of the $U(1)_B$ tachyon, these particles decay to the 
UV sector with the rate determined in Sect.~\ref{adsdecays}.

The approximate $SO(4)$ symmetry of the IR region suppresses decays violating 
the total $SO(4)$ charge. Therefore, a sizeable fraction of the energy 
density in a KS throat is generically in the form of charged KK modes. We 
will now argue that tachyons are also relevant for the decay of these states. 
Namely, the spectrum in~\cite{Ceresole:1999zs} contains tachyons with 
various charges, though not all of them have the maximally allowed negative 
mass-squared. For example, the tachyon in the $SO(4)$ vector multiplet is 
in the adjoint of $SO(4)$. We expect that these tachyons mix with other 
scalars with the same charge in the IR region of the KS throat. Various 
IR-localized states can therefore decay via a given tachyon to the UV 
sector, and the decay rates from Sect.~\ref{adsdecays} apply. As another 
example, we can consider KK modes which are dual to glueballs created by
the operator of lowest dimension, $\smash{\Delta=\frac{3}{2}}$, in the KS 
theory. These states are scalars in the $\smash{(\frac{1}{2},\frac{1}{2})}$
of $SU(2)\times SU(2) \sim SO(4)$. In \cite{Dymarsky:2008wd}, the lightest 
KK mode in this tower was proposed as a candidate for the lightest state
in the KS spectrum. A scalar operator of dimension 
$\smash{\Delta=\frac{3}{2}}$ is dual to a tachyon with 
$\smash{\alpha = 2- \Delta = \frac{1}{2}}$ (cf.~Sect.~\ref{cftpicture}). 
The decay rate of the corresponding KK modes then follows from the 
formulas in Sect.~\ref{adsdecays}.

We now discuss the UV-brane mass term of the various tachyonic scalars. 
As emphasized before, we focus on compactifications where all moduli are 
stabilized by fluxes and non-perturbative effects along the lines of~\cite{
Kachru:2003aw,Giddings:2001yu}. Let us first assume for simplicity that both 
complex structure and K\"ahler moduli are stabilized at the UV scale. 
In 5d language this means that the UV brane theory has only one energy 
scale: $k\sim R^{-1}\sim L^{-1}\sim M_{\rm string}$. (For 
simplicity, we ignore the hierarchy between $L^{-1}$ and $M_{\rm string}$, 
assuming it to be small.) Since the full construction is stable, 
we know that all tachyonic scalars will receive a UV-brane mass term with 
$\lambda\ge\lambda_0$ in this context. Given that there is no small energy 
scale around (which could correspond to $\lambda-\lambda_0\ll 1$) and no 
obvious reason for the tuning $\lambda=\lambda_0$, we make the assumption 
that $\lambda=\lambda_0+{\cal O}(1)$. As we will see in 
Sect.~\ref{cosmology}, even with this conservative assumption the 
cosmological effects of tachyonic scalars can be dramatic.

There are, however, at least two possible loopholes in this conclusion. 
The first loophole is related to the question where the massless mode of a 
5d tachyon with tuned UV-brane mass is localized. To see this, let us first
strengthen our previous argument for detuning in the following 
way: Recall that the wavefunction of the massless mode for supersymmetric 
boundary conditions, $\psi \propto z^{\frac{1}{2} - \alpha}$,
is UV-localized for $\alpha>1$ (cf.~Sects.~\ref{tmt} and 
\ref{cftpicture}). Therefore, even if we change the boundary condition at the
IR brane (e.g. by choosing a different boundary mass term), this mode
only picks up an exponentially small mass. It is then clear that, for
tuned UV mass term and $\alpha>1$, there is always an exponentially
light UV-localized mode in the spectrum. However, with all Calabi-Yau moduli 
stabilized at the high scale, there should, in 5d language, be no light fields 
at or near the UV brane. Thus, the UV-localized mode can not be light, implying 
that the UV mass term has to be detuned. For $\alpha<1$, however, this argument 
does not apply as the wavefunction is IR-localized in this case.
The zero mode of such a tachyon can be light without violating the 
requirement that there are no light fields in the UV sector. Furthermore,
for generic IR-boundary condition, this mode picks up a mass of the order the 
IR scale. Thus, for $\alpha<1$, the UV mass term can in principle be tuned.
Of course, we still have no reason for the required tuning, but we can 
also not dismiss the tuned situation on general grounds.

The second loophole lies in our assumption that there are no light fields
in the UV sector. There is obviously the possibility that the K\"ahler
stabilization scale is much smaller than the flux stabilization scale, which 
may be the natural choice for low-scale SUSY. In this case, the UV-brane 
theory has a second, smaller energy scale with corresponding light fields. 
In the 5d description, these fields could be the lowest KK modes of 
tachyons. More
precisely, for tuned $\lambda = \lambda_0$ and $\alpha>1$, such a mode is massless
and localized in the UV, corresponding to an unstabilized modulus. Stabilization
of the K\"ahler moduli, in the 5d description, is then due to a detuning of
the mass on the UV brane. If the K\"ahler stabilization scale is lower than
the flux stabilization scale (which we assume to be of the order the AdS scale), 
this detuning is small, $\lambda-\lambda_0\ll 1$, giving the modulus a small
mass compared to the AdS scale. Of course, as far as some generic K\"ahler modulus 
is concerned, such a connection is far from obvious because the throat has no 2- or 
4-cycles.\footnote{Note 
that $T^{1,1}$ is isomorphic to $S^2 \times S^3$ and thus has a nontrivial 
2-cycle \cite{Klebanov:1998hh}. However, this $S^2$ shrinks to zero at the 
IR tip of the throat and thus is not a cycle of the Calabi-Yau.
} 
However, such an approximate tuning can potentially be related to the 
universal K\"ahler modulus.\footnote{It 
is clear that the universal K\"ahler modulus belongs to the UV sector if 
we can work in a K\"ahler-Weyl frame (the Brans-Dicke frame). In this 
frame, the metric in the IR region is not affected by a (small) shift of 
the universal K\"ahler modulus. Namely, such a shift corresponds to 
changing the prefactor of the 4d Einstein-Hilbert term, which is dominated 
by the compact Calabi-Yau and the UV end of the throat. However, it has 
recently been argued that this is not always the appropriate 
frame~\cite{Douglas:2008jx,Frey:2008xw}. In other frames, the universal 
K\"ahler modulus may be viewed (at least partially) as a throat field.
} 
If the detuning is sufficiently small, the higher KK modes of the 
corresponding tachyon would decay with the enhanced rates derived in 
Sect.~\ref{tmt}.

In summary, we have very good reasons to consider decays mediated by tachyons 
in throat cosmology. While the situation with detuned UV mass term appears to 
be generic, we can not exclude the possibility that certain tachyonic fields
have a UV mass tuned to the minimal allowed value. Establishing this, which 
would imply even larger decay rates than we discuss in the following, would 
be an interesting subject for future research. 

Finally, we note a recent paper \cite{Frey:2009qb} which presented a survey of 
decay channels of KK modes in a KS throat. In particular, the decay rate of
KK modes to moduli and their axionic partners was determined. As the moduli can be 
viewed (at least partially) as fields living on the UV brane, it is interesting to 
compare this decay rate with our results. More precisely, Eq.~(5.5) in 
\cite{Frey:2009qb} gives the decay rate of a KK mode into axionic partners
of the moduli.\footnote{Note that the decay rate in their 
Eq.~(5.4) is additionally suppressed as it involves a transition between states with 
different charges under the approximately conserved symmetries of the throat.} 
For $M_p \sim M_s \sim k$ and using
our notation $w = m_{\ir} / k$ and $\nu^{\star} = \alpha$, their Eq.~(5.5) reads
\be
\label{comparison}
\Gamma \sim \left(\frac{m_\ir}{k} \right)^{4 + 2 \alpha} m_\ir \, .
\ee
This agrees with our Eq.~\eqref{drs} (apart from the
logarithmic suppression in the case $\alpha = 0$). However, we believe that this 
rate also applies to the tunneling of KK modes to other throats. As discussed in
\cite{Harling:2007jy} in some detail, via the AdS/CFT correspondence, other throats 
can be described by large-$N$ gauge theories which live on the UV brane. The tunneling
rate of KK modes to these sectors is then given by Eq.~\eqref{comparison} times the 
number of degrees of freedom, $\sim N^2$, of the dual gauge theories.

\section{Cosmology}
\label{cosmology}

To illustrate the relevance of tachyons, we will now discuss the 
reheating of the Standard Model (SM)\footnote{We 
use the term `Standard Model' to refer to a sector which contains the SM, 
e.g. the MSSM. For definiteness, we use $g_\sm = \mathcal{O}(100)$ for 
the number of effective relativistic degrees of freedom at the reheating 
temperature $T_{\rh}$. For example, the SM itself has $g_\sm \approx 107$ 
for temperatures $T \gtrsim 300\, \text{GeV}$ while the MSSM has $g_\sm 
\approx 229$ for $T \gtrsim m_{1/2}$ (with the familiar prefactor $7/8$
for fermions included).
} 
after warped brane inflation. For definiteness, we focus on a specific 
realization of warped brane inflation, the KKLMMT scenario 
\cite{Kachru:2003sx}. In this scenario, inflation is driven by a D3-brane 
which slowly rolls towards an anti-D3-brane located at the tip of a KS 
throat. As discussed in Appendix C of \cite{Kachru:2003sx}, for given 
parameters (string coupling, string scale, etc.), the hierarchy of this 
throat can be fixed by observational data.\footnote{Some fine-tuning is 
generically required to actually achieve slow roll inflation. A systematic 
study was carried out in \cite{Baumann:2008kq} (see also \cite{
Baumann:2006th,Burgess:2006cb} for earlier 
work). Some of the mechanisms used in these papers rely on D7-branes 
wrapping 3-cycles of the throat (see e.g. \cite{Chen:2009nk} for an explicit 
realization). For our purposes, we assume a proper fine-tuning such that a 
KKLMMT-type scenario is realized.} In the following, we will use
$m_\ir \sim 10^{14} \, \text{GeV}$ and $k\sim10^{18} \text{ GeV}$ for 
numerical estimates. We assume that the SM is realized on D-branes which 
are localized in the CY outside the throat. In the effective 5d 
description, the SM then lives on the UV brane of the corresponding RS 
model. Furthermore, we assume that the inflationary throat is the only 
strongly warped region in the compact space.

Inflation ends with the annihilation of the D3-brane and the 
anti-D3-brane. The resulting energy is deposited in KK modes localized at 
the tip of the throat \cite{Barnaby:2004kz,Barnaby:2004gg,
Kofman:2005yz,Chialva:2005zy,Frey:2005jk}. Temperature and energy density
of this gas, which is initially marginally non-relativistic, are set by 
the IR scale: $T\sim m_{\ir}$ and $\rho\sim m_{\ir}^4$. These KK modes 
decay to the SM at later times. The resulting reheating temperature of the 
SM can be estimated as usual: Most of the energy is transferred when the 
Hubble rate is comparable to the decay rate. The Hubble rate is related 
to the total energy density by $H^2 = \rho_{\tot} / 3 M_{4}^2$. For 
radiation at temperature $T$ and with $g_\sm$ degrees of freedom, the
energy density is given by $\rho_{\tot}  = \frac{\pi^2}{30} g_\sm T^4$. 
The reheating temperature can then be estimated as
\begin{equation}
\label{reheattemp}
T_{\rh} \sim \left(\frac{10}{g_{\sm}}\right)^{\frac{1}{4}} \, \sqrt{M_{4}
\Gamma} \; .
\end{equation}
It is probable that the energy density in the throat is in the form
of different species of KK modes with different lifetimes. After one species 
of KK modes has decayed to the SM, KK modes with longer lifetimes can easily come 
to dominate the total energy density as their energy density scales like
matter (and not like radiation). Their decay to the SM then leads to a new phase of reheating 
and the final reheating temperature will be determined by the most stable KK modes. 
Let us now assume that the throat has no tachyonic scalars. There certainly are scalar
KK modes in the spectrum of the KS throat (e.g.~from the dilaton). Without tachyons,
their largest possible decay rate would be that for a massless scalar, 
corresponding to $\alpha=2$. 
As discussed in Sect.~\ref{ApplicationstoThroats}, we expect 
that scalars generically obtain a mass on the UV brane (in the 5d description) and that 
this mass is not tuned to the special value $\lambda_0$. Using the corresponding decay rate 
for a massless scalar, Eq.~\eqref{drs}, we then obtain an upper bound on the reheating temperature
in a scenario without tachyons:
\be
T_\rh \, \lesssim \, 1 \text{ GeV} \, .
\ee
This is dangerously low. In particular, it is difficult to obtain sufficient baryogenesis 
at such low temperatures.
Moreover, reheating temperatures below 1 MeV are excluded by nucleosynthesis 
\cite{Kawasaki:2000en}.
As tachyons lead to higher decay rates, the reheating temperature
can also be higher. We will now discuss this in more detail.

We first consider processes in the throat sector which happen on timescales 
shorter than those relevant for decays to the SM. Warped KK modes are described by an effective
field theory with a low cutoff, $\sim m_\ir$. We do not assume a large hierarchy between
$k$ and $M_5$ (or, from the 10d perspective, $k$, $M_s$ and $M_{10}$) which would suppress 
quantum corrections. All possible n-point-interactions, which are 
allowed by the symmetries, will therefore be
induced by quantum effects (if they are not present at tree-level). In particular, the 
effective action includes couplings of the type
\be
\label{4-point}
\mathcal{H}\mathcal{H}\mathcal{G}\mathcal{G},
\ee
which involves two different species of KK modes $\mathcal{H}$ and $\mathcal{G}$
and allows for the processes $2 \cdot \mathcal{H} \leftrightarrow 2 \cdot \mathcal{G}$. 
Here, we have suppressed all symmetry-indices (which are appropriately contracted). 
Let $\mathcal{G}$ denote the lightest KK mode in the spectrum, whereas $\mathcal{H}$ 
is any heavier KK mode. As we will see in a moment, the KK modes are 
in thermal equilibrium initially. For temperatures below the mass of the 
heavier KK mode, the process $2 \cdot \mathcal{H} \rightarrow 2 \cdot \mathcal{G}$ then occurs
with a much higher rate than the inverse process. The heavier KK modes thus
begin to annihilate to the lighter KK modes, leading to an exponential decrease 
of the relative number density of the two states,
\be
\label{suppression}
\frac{n_\mathcal{H}}{n_\mathcal{G}} \, = \, e^{-(m_\mathcal{H}-m_\mathcal{G})/T} \,.
\ee
Here, $m_\mathcal{H}$ and $m_\mathcal{G}$ are the masses of the two 
states and $T$ is the temperature of the gas of KK modes. This exponential decrease continues 
until the heavier
KK modes are so dilute that they decouple. This happens, when
\be
\label{decoupling}
n_\mathcal{H} \cdot \langle \sigma v \rangle \, \sim \, H \, ,
\ee
where $\langle \sigma v \rangle$ is the thermally averaged product of cross section 
and relative velocity for the scattering process and $H$ is the Hubble rate.
The latter is dominated by the energy density $\rho$ of the gas of KK modes, 
$H \sim \sqrt{\rho}/M_4$. Due to the exponential dependence on the 
temperature in Eq.~\eqref{suppression}, 
the heavy KK modes decouple when the temperature is still of the order $m_\ir$.\footnote{More precisely,
the condition on the decoupling temperature $T_\dec$ follows from Eqs.~\eqref{suppression} and \eqref{decoupling}.
Defining $m \equiv m_\mathcal{H}-m_\mathcal{G}$ (and assuming that $m \sim m_\ir$), the
condition can be written as $\smash{e^{-m/T_\dec} \sim  (m/M_4)}$, where we have neglected powers of $m/T_\dec$ 
multiplying the right-hand side. 
For $\smash{m_\ir/M_4 \sim 10^{-4}}$, we 
then find that $T_\dec \sim m_\ir/9$.}
By dimensional analysis, it then follows that $\smash{\langle \sigma v \rangle \sim m_\ir^{-2}}$ 
as well as $\smash{\rho \sim m_\ir^4}$ and
$\smash{n_\mathcal{G} \sim m_\ir^3}$. Using Eq.~\eqref{decoupling}, we find
\be
\frac{n_\mathcal{H}}{n_\mathcal{G}} \, \sim \, \frac{m_\ir}{M_4} \, \sim \, 10^{-4} \, .
\ee
The fact that this factor is smaller than 1 shows that the KK modes were indeed in thermal equilibrium
initially. Thus, the energy density is dominantly in the form of the lightest KK mode. All heavier
KK modes are $10^{-4}$ times less abundant.

There are other decay 
processes in the throat sector. 
In particular, the effective action will also contain various 
trilinear couplings. If kinematically allowed, KK modes 
can therefore decay to two lighter KK modes. If these processes are 
not suppressed due to the approximate symmetry 
$SO(4) \times \mathbb{Z}_2 \times \mathbb{Z}_2$ of the effective action 
(cf.~Sect.~\ref{ApplicationstoThroats}), the corresponding decay
rate is $\sim m_\ir$ (as follows by dimensional analysis). Moreover, 
we expect various couplings of KK modes to 4d gravitons in the effective action.
While such couplings are forbidden by KK-mode orthogonality at the level of
the two-derivative, quadratic action \cite{Chen:2006ni,Dufaux:2008br},
they are certainly present in the strongly-coupled and highly non-linear
effective field theory relevant for the dynamics of the low-lying KK
modes.

Indeed, integrating out the heavier states, all types of multi-particle and
higher-derivative vertices are generated in the effective action for the
lowest-lying KK modes. Since any derivatives appearing in this action
are covariant, various quantities derived from the Riemann tensor
naturally arise. As we assume no significant hierarchy between the AdS scale
and the Planck scale, the only scale relevant for this argument is the
IR scale $m_\ir$. Thus, for example, we naturally expect a term of the
type
\be
\label{EffCoupling}
\sim \frac{1}{m_\ir} R^2 {\cal G}
\ee
to arise. Here, $R^2$ is some scalar quadratic in the 4d Riemann tensor
and the scalar KK mode ${\cal G}$ is a singlet with respect to the
symmetries of the effective action (otherwise, the term 
would be forbidden or suppressed). We therefore believe that such terms are generically 
present in the effective action and that they are not suppressed by powers of the warp 
factor, a possibility raised in \cite{Frey:2009qb}. As discussed in 
\cite{Delbourgo:2000nq,Frey:2009qb}, the coupling Eq.~\eqref{EffCoupling} allows for 
the decay of ${\cal G}$ to two 4d gravitons. By canonically normalizing the gravitons,
the relevant vertex is seen to be suppressed by $1/M_4^2$, leading to the 
decay rate $\sim m_\ir^5/M_4^4$. Similar arguments can be made for couplings involving 
two KK modes and a graviton. Heavier KK modes can therefore
decay to lighter KK modes via the emission of a graviton. If these processes are 
not suppressed due to the approximate symmetries 
of the effective action, the corresponding decay rate is 
$\sim m_\ir^3/M_4^2$.

Let us summarize what we have found: At late times, the energy density is dominantly 
in the form of 
the lightest KK mode. As we have discussed in Sect.~\ref{ApplicationstoThroats},
the lightest known KK mode in the KS throat mixes with the tachyon from the 
Betti multiplet. We will assume that this KK mode is indeed the lightest 
state. It thus decays to the SM with the rate for a maximally 
tachyonic scalar:
\be
\label{drs3}
\Gamma \, \sim \, 
\left(\frac{m_\ir}{k}\right)^4\frac{m_{\ir}}{\ln^2(k/m_\ir)} \, .
\ee
As the lightest known KK mode is odd with respect to the $\mathcal{I}$-symmetry
(cf.~Sect.~\ref{ApplicationstoThroats}), the decay to two gravitons is 
forbidden or strongly suppressed (compared
to the rate $\sim m_\ir^5/k^4$ for $k \sim M_4$). Thus, the decay to the SM is the 
dominant decay channel for this state. 

The heavier KK modes are $\smash{10^{-4}}$ times less abundant than the 
lightest KK mode. In addition, several of these states decay to lighter KK modes via
trilinear couplings and graviton emission.
The timescales for these decays are $\smash{\sim m_\ir^{-1}}$ and $\smash{\sim M_4^2/m_\ir^3}$, respectively. 
At later times, only the lightest KK modes with given charges under 
$SO(4) \times \mathbb{Z}_2 \times \mathbb{Z}_2$
may survive. Moreover, as this symmetry is only approximate, 
eventually all KK modes can decay to the lightest KK mode. Such charge violating
decays are, however, suppressed by additional powers of the warp factor 
\cite{DeWolfe:2004qx,Aharony:2005ez,Berndsen:2008my,Dufaux:2008br}. If the
decay rate is nevertheless higher than Eq.~\eqref{drs3}, all KK modes
will first decay to the lightest KK mode. The latter subsequently decay to the SM. 
Using Eq.~\eqref{reheattemp}, the resulting reheating temperature of the SM is
\be
T_\rh \, \sim \, 10^7 \text{ GeV} \, .
\ee
This is 
considerably higher than 1 GeV, due to the enhanced decay rate mediated by a tachyon. 
It is also possible, however, that the total decay rate of the heavier KK modes 
(to the SM and to the lightest 
KK mode) is smaller than Eq.~\eqref{drs3}. These states will then still
be stable when the lightest KK mode has already decayed to the SM. 
As their energy density scales like matter 
whereas the energy density of the SM scales like radiation, the heavier KK modes might
come to dominate the total energy density. Their decay
to the SM then leads to a new phase of reheating with a lower reheating temperature. 
Although we have not analysed this possibility in more detail, it is clear that tachyons 
may also enhance the relevant decay rates (and thus the reheating temperature) in this case.

Finally, we briefly consider the decay of fermionic KK modes. If supersymmetry is broken
outside the throat, we expect that the fermionic decay rates are of the same order of magnitude as those 
of the scalars in the same multiplet. However, the decays of fermions are somewhat model-dependent 
as the supersymmetry breaking scale (or the gravitino mass $m_{3/2}$) determines the allowed 
decay channels. If supersymmetry is broken at low scales, $m_{3/2} < m_\ir$\footnote{In some scenarios 
of brane inflation the gravitino 
mass provides an upper bound on the inflationary Hubble scale, $H_I \lesssim  m_{3/2}$ 
\cite{Kallosh:2004yh,Conlon:2008cj}. Since 
$H_I \sim m_{\ir}^{2} / M_{4}$, the case $m_{3/2} < m_\ir$ is possible also in these scenarios.}, the 
fermionic KK modes can decay to lighter KK modes under the emission of a gravitino. Another possible 
channel is the decay to a graviton and a gravitino. These processes are the analogue of the decay
of bosonic KK modes to gravitons. The resulting abundance of gravitinos leads to the well-known gravitino problem.
If the supersymmetry breaking scale is high, $m_{3/2} > m_\ir$, decays to gravitinos and decays to 
superpartners of standard model particles are kinematically forbidden. As we have assumed that the
inflationary throat is the only strongly warped region (and the fermionic KK modes can thus not decay
to such sectors), the fermionic KK modes may then be absolutely
stable. In order to avoid the resulting overclosure of the universe, one can invoke the neutrino portal 
which allows their decay to a neutrino and a Higgs \cite{Harling:2008px,Falkowski:2009yz}.

\section{Conclusions}
\label{conclusions}

In this paper, we have analysed the effect of tachyonic scalars on 
couplings between IR- and UV-localized sectors in warped compactifications. 
We gave an introduction to tachyonic scalars in a slice of AdS$_5$ in 
Sect.~\ref{AdS5fields}. In particular, we explained the origin of an 
instability and its removal via a UV-brane-localized mass term for the field, 
both of which can be easily understood in a quantum mechanical analogue. 
In Sect.~\ref{adsdecays} we considered decays of IR-localized KK-modes to 
a UV-localized gauge theory in a 5d Randall-Sundrum toy model and derived 
the dependence of the decay rates on the warp factor. As expected, this 
dependence is governed by the 5d mass of the scalar. There is also a 
further suppression due to the UV-localized mass term.

We have developed the dual CFT description of our results in 
Sect.~\ref{cftpicture}. In this approach, the dependence on the 5d mass
$M$ is encoded in the dimension of the dual 
operator, given by $\Delta=2+\sqrt{4+M^2/k^2}$ . As long as the UV-brane 
mass term takes generic values 
above the boundary for stability, its effect on decay rates can be
understood as a simple propagator suppression on the CFT side. If this
boundary mass term is tuned to its minimal allowed value, decay rates are 
correspondingly enhanced. While this is straightforward to see for 
$\Delta>3$, the analysis of such a tuned situation in the regime with 
$\Delta<3$ is more subtle: Here, one has to use the alternative CFT 
description with an operator of dimension $\Delta' = 4 - \Delta$. 

In Sect.~\ref{ApplicationstoThroats}, we have worked out the applicability
of our 5d analysis to throat geometries in string compactifications. 
As a specific example, we have considered the Klebanov-Strassler throat 
with its approximate AdS$_5 \times T^{1,1}$ geometry in the UV region. Due 
to the (gauge) symmetries of the solution, tachyonic 5d scalars saturating 
the Breitenlohner-Freedman bound are present. We have argued that, 
generically, the UV-brane mass of these tachyons is detuned from its
minimal value.\footnote{We 
have, however, also pointed out potential loopholes in this argument. 
Thus, the case of a tuned UV mass term can not be completely dismissed. 
Such a tuning would allow for decay rates which are even higher than those 
which we found in the generic situation.
} 
The spectrum of light KK modes in the 
Klebanov-Strassler throat has been studied (cf. e.g. \cite{Benna:2007mb,
Dymarsky:2008wd,Krasnitz:2000ir,Caceres:2000qe,Amador:2004pz,Noguchi:2005ws,
Berg:2006xy,Dymarsky:2007zs}) and the lightest state found so far is a singlet under 
the continuous symmetries of the throat. This state results from the 
mixture of two 5d scalars, one of which has maximally tachyonic 5d mass. 
Using the equations of motion of the two coupled scalars, we have checked 
that the two fluctuations decouple in the UV and that one of the resulting 
wavefunctions correctly describes a maximally tachyonic scalar. The 
lightest (known) KK mode of the Klebanov-Strassler throat thus decays to 
the UV brane with the decay rate derived in Sect.~\ref{adsdecays}. 

In Sect.~\ref{cosmology}, we have used a specific example, the reheating of 
the standard model after warped brane inflation, to demonstrate the relevance 
of tachyons in throat cosmology. Since tachyons affect the decay 
rate to the UV brane, their role is decisive in settings where the SM is 
localized outside the inflationary throat. We have focused on such scenarios 
assuming, for definiteness, that the SM is realized in the bulk of the 
Calabi-Yau.\footnote{We 
believe, however, that our results also apply to situations where the SM 
lives in another throat. In this case, our decay rates have to be multiplied 
by the number of degrees of freedom, $\sim N^2$, of the large-$N$ gauge 
theory which is dual to the additional throat.
}
An important consequence of tachyons is that they lead to a 
larger reheating temperature of the standard model. More precisely, 
without tachyons, one finds an upper bound on the reheating temperature 
$T_\rh \, \lesssim \, 1 \text{ GeV}$. Including the tachyons, 
one instead has $T_\rh \, \sim \, 10^7 \text{ GeV}$. Charged KK modes 
may have smaller decay rates and correspondingly larger lifetimes. Their 
decay may lead to another phase of reheating, resulting in a lower 
reheating temperature. Although we have not determined the decay rate of 
a generic charged KK mode, it is likely that their decay rates are 
also enhanced by tachyons.

To discuss this in more detail, we recall that the UV region of the 
Klebanov-Strassler throat can be approximated as 
AdS$_5\times T^{1,1}$. In this region, the KK reduction on $T^{1,1}$, 
which was performed in \cite{Ceresole:1999zs}, is applicable. Close to
the IR end of the
throat, the approximation as AdS$_5 \times T^{1,1}$ becomes unreliable.
We expect that a given AdS$_5$ field (coming from the KK reduction on 
$T^{1,1}$) mixes with all other fields with the same quantum numbers in this region.
This is likely already the case at tree-level as follows e.g.~from the analysis
in \cite{Berg:2006xy,Benna:2007mb,Dymarsky:2008wd}. Moreover,
there generically is no large hierarchy between the 10-dimensional Planck scale 
and the AdS curvature scale which could strongly suppress quantum corrections.
Thus, we expect all bilinear mixing terms, which are allowed by the symmetries,
to appear in the effective 5d action. A 4d KK mode with given quantum numbers
will therefore couple to the UV-brane via all AdS$_5$ fields with the same
quantum numbers. It is then clear that the decay rate of a scalar KK mode
will be determined by the AdS$_5$ field with the smallest (or possibly most
tachyonic) mass-squared for the given charges.  As the spectrum of these
fields is known \cite{Ceresole:1999zs}, it should be possible to 
determine the decay rate of a generic scalar KK mode from its
charges under the (approximate) symmetries of the throat.
This would be an interesting project for future research.

Another important cosmological application of our results is to 
throat dark matter 
\cite{Kofman:2005yz,Chen:2006ni,Berndsen:2008my,Dufaux:2008br,Harling:2008px,Chen:2009iua,Frey:2009qb}.
Let us first consider scenarios in which the standard model is 
localized outside the throat. Measurements of the cosmic diffuse $\gamma$-ray
background give a lower bound of $\smash{\sim 10^{26}}$ s for the lifetime
of dark matter with decay channels to photons (assuming an $\mathcal{O}(1)$ branching 
ratio for decays via hadrons) \cite{Kribs:1996ac}. 
As we have discussed in
Sect.~\ref{ApplicationstoThroats}, the lightest known KK mode of 
the Klebanov-Strassler throat couples to the UV-brane via a maximally tachyonic scalar.
Generically, a considerable fraction of the dark matter will be in the form of this 
lightest state. 
In order to fulfill the above bound, the dark matter throat then has 
to have an IR scale which is smaller than $\sim 10^5$ GeV (assuming ${k \sim 10^{18} \text{ GeV}}$).
This effectively excludes KK modes in an inflationary throat \`a la KKLMMT \cite{Kachru:2003sx} 
as a viable dark matter candidate. Dark matter in sufficiently long throats, which may be 
produced by tunneling from the inflationary throat \cite{Kofman:2005yz,Chen:2006ni} or thermally from the standard 
model \cite{Harling:2008px}, can however still be viable.
It may, of course, turn out that the lightest
KK mode of the Klebanov-Strassler throat is not the one considered in
Sect.~\ref{ApplicationstoThroats}. However, the mass of a glueball can 
roughly be expected to decrease with the dimension of the operator which 
creates the state \cite{Benna:2007mb,Dymarsky:2008wd}. Or, correspondingly,
the mass of KK modes generically decreases with the mass of the corresponding
AdS$_5$ field \cite{Berndsen:2008my}. It is therefore likely that 
any lighter KK mode is also associated with a 5d tachyon which determines
its decay rate. Finally, dark matter can be formed by fermionic KK modes
which can decay to the standard model via the neutrino portal 
\cite{Harling:2008px,Falkowski:2009yz}. The decay rate is likely to be determined by the
superpartner of a tachyonic 5d scalar and can be suppressed by (approximate) R-parity. 

Alternatively, in scenarios in which the standard model is located in the dark matter throat
\cite{Kofman:2005yz,Chen:2006ni,Berndsen:2008my,Dufaux:2008br,Frey:2009qb},
our results are relevant for the decay of dark matter to other throats (which are 
likely to be present in a given compactification \cite{Hebecker:2006bn}). If the 
dark matter consists of the lightest (known) KK mode\footnote{Note that this state is odd under a 
$\mathbb{Z}_2$-symmetry and may therefore be stable against decay to the standard model.},
the requirement that the dark matter lives longer than the current age of the universe gives an 
upper bound of $ \sim 10^7$ GeV on the IR scale of the throat (again assuming $k \sim 10^{18}$ GeV). 
The preferred scale $\sim$ TeV thus
still allows for a viable dark matter candidate (provided it is sufficiently stable against decay 
to the standard model). As we have discussed above, even if the 
dark matter instead consists of other charged KK modes, tachyons are again likely to determine 
their decay rate to the UV sector.

\subsection*{Acknowledgements}
We would like to thank Johanna Erdmenger, Tony Gherghetta, Dam T. Son and Gianmassimo 
Tasinato for helpful comments and discussions. A.H. is grateful to 
the Berkeley Center for Theoretical Physics for hospitality. This 
work was supported by the German Research Foundation (DFG) within
the Transregional Collaborative Research Centre TR33 ``The Dark 
Universe''.

\section*{Appendix: Decay Rates with Bessel Functions}
\label{BesselCalc}

The purpose of this appendix is to compute the decay rates estimated  
in Sect.~\ref{adsdecays} explicitly in terms of Bessel functions.  
Similar calculations to those below can be found e.g. in
\cite{Gherghetta:2000qt,Davoudiasl:1999tf,Chung:2000rg,Dimopoulos: 
2001ui}.

The analytic solution to Eq.~\eqref{schroedinger2} in terms of Bessel  
functions is given by
\begin{equation}
\label{Besselsol}
\psi_{n} = \frac{1}{a_{\alpha,n}} \widehat{z}^{\, \, 1/2} \,  
\left[ J_{\alpha}(\widehat{z}) + b_{\alpha,n}  \, Y_{\alpha} 
(\widehat{z})\right] \, ,
\end{equation}
where $a_{\alpha,n}$ and $b_{\alpha,n}$ are constants determined from  
the normalization of $\psi_n$, Eq.~\eqref{normalization}, and the  
boundary conditions, Eq.~\eqref{fullBCz}.

In order to relate the analytic solution to the previous estimates,  
we consider the asymptotic expansions of the Bessel functions in the  
regions of small and large argument \cite{Abramowitz}. For $0 < z \ll  
\sqrt{\alpha + 1}$:
\begin{equation}
\label{Jalphasmallz}
J_{\alpha}(z) \, \simeq \,\displaystyle \frac{1}{\Gamma(\alpha + 1)} \,
\left(\frac{z}{2}\right)^{\alpha} 
\end{equation}
and
\begin{equation}
\label{Yalphasmallz}
Y_{\alpha}(z) \, \simeq \, \displaystyle
\begin{cases}
   \, \frac{2}{\pi} (\gamma - \ln 2 + \ln z) & \text{for} \; \alpha =  
0 \\
   & \\
   \displaystyle \, - \frac{\Gamma(\alpha)}{\pi} \, \left(\frac{2}{z} 
\right)^{\alpha}
-\frac{\cos(\pi \alpha) \Gamma(-\alpha)}{\pi} \left(\frac{z} 
{2}\right)^{\alpha} &
\text{for} \; \alpha > 0 \, .
\end{cases}
\end{equation}
Here, $\gamma \simeq 0.5772$ is the Euler constant and $ 
\Gamma(\alpha)$ is
the gamma function. The second term in the expansion for $Y_\alpha$ is  
important in the limit $\alpha \rightarrow 0$. Furthermore, for $z \gg  
\left|\alpha^{2} - \frac{1}{4}\right|$:
\begin{equation}
\label{Jalphalargez}
J_{\alpha}(z) \simeq \sqrt{\frac{2}{\pi z}} \cos \left(z -  
\frac{\alpha \pi}{2}
- \frac{\pi}{4}\right) 
\end{equation}
and
\begin{equation}
\label{Yalphalargez}
Y_{\alpha}(z) \simeq \sqrt{\frac{2}{\pi z}} \sin \left(z -  
\frac{\alpha \pi}{2}
- \frac{\pi}{4}\right) \, .
\end{equation}
From these expansions it is clear that the Bessel functions  
explicitly realize the matching of UV and IR solutions discussed in  
Sect.~\ref{adsdecays}, cf. Eqs.~\eqref{smallzsol} and \eqref{largez}.

Moreover, the Bessel functions satisfy certain identities which are  
useful in the calculation of the decay rates below. For any linear  
combination $\mathcal{C}_{\alpha}(z)$ of Bessel functions $J_{\alpha} 
(z)$ and $Y_{\alpha}(z)$ in which the coefficients are independent of $ 
\alpha$ and $z$,
the following identities are valid \cite{Abramowitz}:
\begin{equation}
\frac{2 \alpha}{z} \mathcal{C}_{\alpha}(z) =
\mathcal{C}_{\alpha + 1}(z) +
\mathcal{C}_{\alpha - 1}(z) \, ,
\end{equation}
\begin{equation}
2 \frac{d}{dz} \mathcal{C}_{\alpha}(z) = \mathcal{C}_{\alpha - 1}(z) -
\mathcal{C}_{\alpha + 1}(z) 
\end{equation}
and
\begin{equation}
\int_{a}^{b} dz \, z \, \mathcal{C}_{\alpha}^{2} = \frac{z^{2}}{2}  
\left[
\mathcal{C}_{\alpha}^{2}(z) - \mathcal{C}_{\alpha + 1}(z)  
\mathcal{C}_{\alpha -
1}(z) \right] \vert_{a}^{b}
\, .
\end{equation}
In addition, for integer $\alpha$, $\mathcal{C}_{\alpha}$ satisfies:
\begin{equation}
\mathcal{C}_{-\alpha}(z) = (-1)^{\alpha} \mathcal{C}_{\alpha}(z) \, .
\end{equation}
Furthermore, the following relation holds for all $\alpha$ and $z$:
\begin{equation}
J_{\alpha + 1}(z) Y_{\alpha}(z) - J_{\alpha}(z) Y_{\alpha + 1}(z) =
\frac{2}{\pi z} \, .
\end{equation}
We will use these identities frequently throughout the computations to  
simplify the results.

By inserting the solution Eq.~\eqref{Besselsol} into the boundary
conditions Eq.~\eqref{fullBCz}, we find:
\begin{equation}
\label{BesselBalpha}
- b_{\alpha,n} = \frac{\widehat{z}_{\ir} J_{\alpha - 1} 
(\widehat{z}_{\ir}) +
(2 - \alpha - \lambda) J_{\alpha}(\widehat{z}_{\ir})}{\widehat{z}_{\ir}
Y_{\alpha - 1}(\widehat{z}_{\ir}) + (2 - \alpha - \lambda) Y_{\alpha} 
(\widehat{z}_{\ir})}
= \frac{\widehat{z}_{\uv} J_{\alpha - 1}(\widehat{z}_{\uv}) + (2 -  
\alpha -
\lambda) J_{\alpha}(\widehat{z}_{\uv})}{\widehat{z}_{\uv}
Y_{\alpha - 1}(\widehat{z}_{\uv}) + (2 - \alpha - \lambda)
Y_{\alpha}(\widehat{z}_{\uv})} \, .
\end{equation}
From Eq.~\eqref{BesselBalpha} we see that the masses $m_n$ are  
determined by
\begin{equation}
\label{massquant}
\left[ \widehat{z}_{\ir} J_{\alpha - 1}(\widehat{z}_{\ir}) +
(2 - \alpha - \lambda) J_{\alpha}(\widehat{z}_{\ir}) \right] +  
b_{\alpha,n} \, \left[ \widehat{z}_{\ir}
Y_{\alpha - 1}(\widehat{z}_{\ir}) + (2 - \alpha - \lambda) Y_{\alpha} 
(\widehat{z}_{\ir}) \right] = 0 \, .
\end{equation}
Using the asymptotic expansions for small arguments, 
Eqs.~\eqref{Jalphasmallz} and \eqref{Yalphasmallz}, as well as 
Eq.~\eqref{BesselBalpha}, we find that (up to $\mathcal{O}(1)$ prefactors)
\begin{equation}
\label{BesselBapprox}
b_{\alpha,n} \sim \begin{cases}
   \, \widehat{z}_{\uv}^{\, \, 2 \alpha} & \text{for generic} \; \lambda \\
   & \\
   \alpha \, (1-\alpha) \, \widehat{z}_{\uv}^{\, \, 2 \alpha - 2} & \text{for} \;  
\lambda = \lambda_0 \, . \\
\end{cases}
\end{equation}
For $k e^{- k\ell} \ll m_n \ll k$ and generic $\alpha$ and $\lambda$, one 
then finds that the masses $m_n$ are  
determined by the zeros of $J_{\alpha-1}(\widehat{z}_{\ir})$. This gives (see e.g. \cite{Gherghetta:2000qt}):
\begin{equation}
\label{genmassivespectrum}
m_n \simeq \left(n + \frac{\alpha}{2} - \frac{3}{4}\right) \pi k e^{-  
k \ell} \, .
\end{equation}
As can be seen from the second line of Eq.~\eqref{BesselBapprox}, the masses are also
determined by the zeros of $J_{\alpha-1}(\widehat{z}_{\ir})$ for a tuned UV mass term,
as long as $\alpha > 1$.
For $\alpha < 1$, the mass spectrum is instead given by the zeros of  
$Y_{\alpha-1}(\widehat{z}_{\ir})$. However, this only leads to  
a shift in the spectrum.

Next, we want to verify the approximate results for the decay rates,
Eqs.~\eqref{drs}, \eqref{0a1} and \eqref{a12}. Defining 
$Z_{\alpha}(z) \equiv J_{\alpha}(z) +
b_{\alpha,n} Y_{\alpha}(z)$, with $b_{\alpha,n}$ given by
Eq.~\eqref{BesselBalpha}, the boundary conditions can be rewritten as
\begin{equation}
\label{boundcondonZalpha}
Z_{\alpha - 1}(\widehat{z}_{\ir}) = - \frac{2 - \alpha - \lambda} 
{\widehat{z}_{\ir}}
Z_{\alpha}(\widehat{z}_{\ir}) \quad \text{and} \quad Z_{\alpha -
1}(\widehat{z}_{\uv}) = - \frac{2 - \alpha - \lambda}{\widehat{z}_{\uv}}
Z_{\alpha}(\widehat{z}_{\uv}) \, .
\end{equation}
Furthermore, we have to determine the normalization constant  
$a_{\alpha,n}$ from
Eq.~\eqref{normalization}. Using the above identities for Bessel  
functions and
the boundary conditions Eq.~\eqref{boundcondonZalpha}, we find
\begin{equation}
\label{Besselnormalization}
a_{\alpha,n} = \displaystyle \frac{1}{m_n} \, \sqrt{  (\widehat{z}_{\ir}^{2} +  
c) Z_{\alpha}^{2}(\widehat{z}_{\ir}) - (\widehat{z}_{\uv}^{2} + c)  Z_{\alpha}^{2} 
(\widehat{z}_{\uv}) } \, ,
\end{equation}
where $c = 2 \alpha \, (2-\alpha-\lambda) + (2-\alpha-\lambda)^2 \sim  
\mathcal{O}(1)$ for generic $\lambda$.

From the definition of the coupling constant, Eq.~\eqref{dimfulcouplingUV}, and
our result for the normalization constant, Eq.~\eqref{Besselnormalization}, we see  
that we have to compute the ratio of $Z_{\alpha}$ evaluated at the IR  
brane to
$Z_{\alpha}$ evaluated at the UV brane. By repeatedly using the boundary  
conditions
Eq.~\eqref{boundcondonZalpha}, as well as Eq.~\eqref{BesselBalpha} and  
the
identities given above, this ratio can be simplified to
\begin{equation}
\label{Zalpharatio}
\frac{Z_{\alpha}(\widehat{z}_{\ir})}{Z_{\alpha}(\widehat{z}_{\uv})} =
\frac{\widehat{z}_{\uv} Y_{\alpha - 1}(\widehat{z}_{\uv}) + (2 -  
\alpha -
\lambda) Y_{\alpha}(\widehat{z}_{\uv})}{\widehat{z}_{\ir} Y_{\alpha -
1}(\widehat{z}_{\ir}) + (2 - \alpha-\lambda) Y_{\alpha} 
(\widehat{z}_{\ir})} \, .
\end{equation}
Since we assume that $\widehat{z}_{\uv} \ll 1$, we can use the small  
argument expansion, Eq.~\eqref{Yalphasmallz}, to simplify the nominator.  
Furthermore, assuming that $\widehat{z}_{\ir} \gg 1$, we obtain the  
following approximation (again up to $\mathcal{O}(1)$  
prefactors) for the denominator using the mass quantization  
condition Eq.~\eqref{massquant}:
\begin{equation}
\label{ratiodenominator}
\widehat{z}_{\ir} Y_{\alpha -
1}(\widehat{z}_{\ir}) + (2 - \alpha-\lambda) Y_{\alpha} 
(\widehat{z}_{\ir}) \sim
\begin{cases}
   \, \displaystyle \sqrt{\widehat{z}_{\ir}} &  \text{for generic}  
\quad \alpha \quad \text{and} \quad \lambda \\
   & \\
   \, \displaystyle \sqrt{\widehat{z}_{\ir}} & \text{for} \quad \alpha  
 > 1 \quad \text{and} \quad \lambda = \lambda_0 \\
   & \\
   \, \displaystyle \sqrt{\widehat{z}_{\ir}} \, \widehat{z}_{\uv}^{\, \, 2 - 2 \alpha} & \text{for} \quad \alpha < 1 \quad \text{and} \quad  
\lambda = \lambda_0 \; .
\end{cases}
\end{equation}
Combining the resulting simplified expression for Eq.~\eqref{Zalpharatio} with
Eqs.~\eqref{dimfulcouplingUV} and \eqref{Besselnormalization}, we can estimate the decay rates on  
dimensional grounds:
\begin{equation}
\label{Besseldecayrate}
\Gamma_n \sim \begin{cases}
   \, \displaystyle \left( \frac{m_n}{k} \right)^{4 + 2 \alpha} m_\ir  
& \text{for generic} \quad \alpha > 0 \quad \text{and} \quad \lambda \\
   & \\
   \, \displaystyle \left(\frac{m_n}{k} \right)^{4} \frac{m_\ir}{\ln^{2}(k / m_n)} & \text{for} \quad \alpha = 0 \quad \text{and generic} \quad \lambda \\
   & \\
   \, \displaystyle \left( \frac{m_n}{k} \right)^{2 \alpha} m_\ir &  
\text{for} \quad \alpha > 1 \quad \text{and} \quad \lambda = \lambda_0  
\\
   & \\
   \, \displaystyle \left( \frac{m_n}{k} \right)^{4 - 2 \alpha} m_\ir  
& \text{for} \quad \alpha < 1 \quad \text{and} \quad \lambda =  
\lambda_0 \;. 
\end{cases}
\end{equation}
Here, we have replaced $\widehat{z}_{\uv}$ by $m_n / k$ and $ 
\widehat{z}_{\ir}$ by $m_n / m_\ir$.
The logarithmic factor for $\alpha = 0$ and generic $\lambda$ 
arises from the logarithm in the expansion of $Y_0$ for small argument, Eq.~\eqref{Yalphasmallz}, 
inserted into Eq.~\eqref{Zalpharatio}. The decay rates in Eq.~\eqref{Besseldecayrate} agree
with the results of our simplified calculation, Eqs.~\eqref{drs}, 
\eqref{0a1} and \eqref{a12}.


\begin{thebibliography}{10}

\bibitem{Giddings:2001yu}
  S.~B.~Giddings, S.~Kachru and J.~Polchinski,
  ``Hierarchies from fluxes in string compactifications,''
  Phys.\ Rev.\  D {\bf 66} (2002) 106006
  [arXiv:hep-th/0105097].

\bibitem{Denef:2004ze}
  F.~Denef and M.~R.~Douglas,
  ``Distributions of flux vacua,''
  JHEP {\bf 0405} (2004) 072
  [arXiv:hep-th/0404116]; \\
  A.~Giryavets, S.~Kachru and P.~K.~Tripathy,
  ``On the taxonomy of flux vacua,''
  JHEP {\bf 0408} (2004) 002
  [arXiv:hep-th/0404243]; \\
  J.~P.~Conlon and F.~Quevedo,
  ``On the explicit construction and statistics of Calabi-Yau flux vacua,''
  JHEP {\bf 0410} (2004) 039
  [arXiv:hep-th/0409215]; \\
  S.~Ashok and M.~R.~Douglas,
  ``Counting flux vacua,''
  JHEP {\bf 0401} (2004) 060
  [arXiv:hep-th/0307049]; \\
  T.~Eguchi and Y.~Tachikawa,
  ``Distribution of flux vacua around singular points in Calabi-Yau moduli
  space,''
  JHEP {\bf 0601} (2006) 100
  [arXiv:hep-th/0510061].

\bibitem{Hebecker:2006bn}
  A.~Hebecker and J.~March-Russell,
  ``The ubiquitous throat,''
  Nucl.\ Phys.\  B {\bf 781} (2007) 99
  [arXiv:hep-th/0607120].

\bibitem{Klebanov:2000hb}
  I.~R.~Klebanov and M.~J.~Strassler,
  ``Supergravity and a confining gauge theory: Duality cascades and
  chiSB-resolution of naked singularities,''
  JHEP {\bf 0008} (2000) 052
  [arXiv:hep-th/0007191].

\bibitem{Brummer:2005sh}
  F.~Brummer, A.~Hebecker and E.~Trincherini,
  ``The throat as a Randall-Sundrum model with Goldberger-Wise
  stabilization,''
  Nucl.\ Phys.\  B {\bf 738}, 283 (2006)
  [arXiv: hep-th/0510113].

\bibitem{Kachru:2003aw}
  S.~Kachru, R.~Kallosh, A.~D.~Linde and S.~P.~Trivedi,
  ``De Sitter vacua in string theory,''
  Phys.\ Rev.\  D {\bf 68}, 046005 (2003)
  [arXiv:hep-th/0301240].

\bibitem{Kachru:2003sx}
  S.~Kachru, R.~Kallosh, A.~D.~Linde, J.~M.~Maldacena, L.~P.~McAllister and S.~P.~Trivedi,
  ``Towards inflation in string theory,''
  JCAP {\bf 0310} (2003) 013
  [arXiv:hep-th/0308055].

\bibitem{Dimopoulos:2001ui}
  S.~Dimopoulos, S.~Kachru, N.~Kaloper, A.~E.~Lawrence and E.~Silverstein,
  ``Small numbers from tunneling between brane throats,''
  Phys.\ Rev.\  D {\bf 64}, 121702 (2001)
  [arXiv:hep-th/0104239] and 
  ``Generating small numbers by tunneling in multi-throat  compactifications,''
  Int.\ J.\ Mod.\ Phys.\  A {\bf 19}, 2657 (2004)
  [arXiv:hep-th/0106128].

\bibitem{Chen:2006ni}
  X.~Chen and S.~H.~Tye,
  ``Heating in brane inflation and hidden dark matter,''
  JCAP {\bf 0606} (2006) 011
  [arXiv:hep-th/0602136].

\bibitem{Berndsen:2008my}
  A.~Berndsen, J.~M.~Cline and H.~Stoica,
  ``Kaluza-Klein relics from warped reheating,''
  Phys.\ Rev.\  D {\bf 77} (2008) 123522
  [arXiv:0710.1299 [hep-th]].

\bibitem{Dufaux:2008br}
  J.~F.~Dufaux, L.~Kofman and M.~Peloso,
  ``Dangerous Angular KK/Glueball Relics in String Theory Cosmology,''
  Phys.\ Rev.\  D {\bf 78} (2008) 023520
  [arXiv:0802.2958 [hep-th]].

\bibitem{Harling:2008px}
  B.~v.~Harling and A.~Hebecker,
  ``Sequestered Dark Matter,''
  JHEP {\bf 0805} (2008) 031
  [arXiv:0801.4015 [hep-ph]].

\bibitem{Chen:2009iua}
  X.~Chen,
  ``Decaying Hidden Dark Matter in Warped Compactification,''
  JCAP {\bf 0909}, 029 (2009)
  [arXiv:0902.0008 [hep-ph]].

\bibitem{Frey:2009qb}
  A.~R.~Frey, R.~J.~Danos and J.~M.~Cline,
  ``Warped Kaluza-Klein Dark Matter,''
  JHEP {\bf 0911} (2009) 102
  [arXiv:0908.1387 [hep-th]].

\bibitem{Barnaby:2004gg}
  N.~Barnaby, C.~P.~Burgess and J.~M.~Cline,
  ``Warped reheating in brane-antibrane inflation,''
  JCAP {\bf 0504} (2005) 007
  [arXiv:hep-th/0412040].

\bibitem{Kofman:2005yz}
  L.~Kofman and P.~Yi,
  ``Reheating the universe after string theory inflation,''
  Phys.\ Rev.\  D {\bf 72} (2005) 106001
  [arXiv:hep-th/0507257].

\bibitem{Chialva:2005zy}
  D.~Chialva, G.~Shiu and B.~Underwood,
  ``Warped reheating in multi-throat brane inflation,''
  JHEP {\bf 0601} (2006) 014
  [arXiv:hep-th/0508229].

\bibitem{Frey:2005jk}
  A.~R.~Frey, A.~Mazumdar and R.~C.~Myers,
  ``Stringy effects during inflation and reheating,''
  Phys.\ Rev.\  D {\bf 73} (2006) 026003
  [arXiv:hep-th/0508139].

\bibitem{Panda:2009ji}
  S.~Panda, M.~Sami and I.~Thongkool,
  ``Reheating the D-brane universe via instant preheating,''
  arXiv:0905.2284 [hep-th].

\bibitem{Randall:1998uk}
  L.~Randall and R.~Sundrum,
  ``Out of this world supersymmetry breaking,''
  Nucl.\ Phys.\  B {\bf 557} (1999) 79
  [arXiv:hep-th/9810155]; \\
  A.~Anisimov, M.~Dine, M.~Graesser and S.~D.~Thomas,
  ``Brane world SUSY breaking,''
  Phys.\ Rev.\  D {\bf 65} (2002) 105011
  [arXiv:hep-th/0111235] and 
  ``Brane world SUSY breaking from string/M theory,''
  JHEP {\bf 0203} (2002) 036
  [arXiv:hep-th/0201256]; \\
  T.~Gregoire, R.~Rattazzi, C.~A.~Scrucca, A.~Strumia and E.~Trincherini,
  ``Gravitational quantum corrections in warped supersymmetric brane  
  worlds,''
  Nucl.\ Phys.\  B {\bf 720} (2005) 3
  [arXiv:hep-th/0411216]; \\
  T.~Gregoire, R.~Rattazzi and C.~A.~Scrucca,
  ``D-type supersymmetry breaking and brane-to-brane gravity mediation,''
  Phys.\ Lett.\  B {\bf 624} (2005) 260
  [arXiv:hep-ph/0505126].

\bibitem{Shuster:1999zf}
  E.~Shuster,
  ``Killing spinors and supersymmetry on AdS,''
  Nucl.\ Phys.\  B {\bf 554} (1999) 198
  [arXiv:hep-th/9902129].

\bibitem{Gherghetta:2000qt}
  T.~Gherghetta and A.~Pomarol,
  ``Bulk fields and supersymmetry in a slice of AdS,''
  Nucl.\ Phys.\  B {\bf 586} (2000) 141
  [arXiv:hep-ph/0003129].

\bibitem{Ceresole:1999zs}
  A.~Ceresole, G.~Dall'Agata, R.~D'Auria and S.~Ferrara,
  ``Spectrum of type IIB supergravity on AdS(5) x T(11): Predictions on N  = 1
  SCFT's,''
  Phys.\ Rev.\  D {\bf 61} (2000) 066001
  [arXiv:hep-th/9905226], \\
  A.~Ceresole, G.~Dall'Agata and R.~D'Auria,
  ``KK spectroscopy of type IIB supergravity on AdS(5) x T(11),''
  JHEP {\bf 9911} (1999) 009
  [arXiv:hep-th/9907216], \\
  A.~Ceresole, G.~Dall'Agata, R.~D'Auria and S.~Ferrara,
  ``Superconformal field theories from IIB spectroscopy on AdS(5) x T(11),''
  Class.\ Quant.\ Grav.\  {\bf 17} (2000) 1017
  [arXiv:hep-th/9910066].

\bibitem{Breitenlohner:1982jf}
  P.~Breitenlohner and D.~Z.~Freedman,
  ``Stability In Gauged Extended Supergravity,''
  Annals Phys.\  {\bf 144} (1982) 249 and
  ``Positive Energy In Anti-De Sitter Backgrounds And Gauged Extended
  Supergravity,''
  Phys.\ Lett.\  B {\bf 115} (1982) 197.

\bibitem{Ghoroku:2001pi}
  K.~Ghoroku and A.~Nakamura,
  ``Stability of Randall-Sundrum brane-world and tachyonic scalar,''
  Phys.\ Rev.\  D {\bf 64} (2001) 084028
  [arXiv:hep-th/0103071].

\bibitem{Delgado:2003tx}
  A.~Delgado and M.~Redi,
  ``Tachyons in a slice of AdS,''
  Phys.\ Lett.\  B {\bf 562} (2003) 127
  [arXiv:hep-th/0301151].

\bibitem{Kaplan:2009kr}
  D.~B.~Kaplan, J.~W.~Lee, D.~T.~Son and M.~A.~Stephanov,
  ``Conformality Lost,''
  arXiv:0905.4752 [hep-th].

\bibitem{Toharia:2008ug}
  M.~Toharia,
  ``Odd Tachyons in Compact Extra Dimensions,''
  arXiv:0803.2503 [hep-th].

\bibitem{Sundrum:2009ii}
 R.~Sundrum and C.~M.~Wells,
 ``Warped Hybrid Inflation,''
 arXiv:0909.3254 [hep-ph].

\bibitem{Randall:1999ee}
  L.~Randall and R.~Sundrum,
  ``A large mass hierarchy from a small extra dimension,''
  Phys.\ Rev.\ Lett.\  {\bf 83} (1999) 3370
  [arXiv:hep-ph/9905221].

\bibitem{Goldberger:1999uk}
  W.~D.~Goldberger and M.~B.~Wise,
  ``Modulus stabilization with bulk fields,''
  Phys.\ Rev.\ Lett.\  {\bf 83} (1999) 4922
  [arXiv:hep-ph/9907447].

\bibitem{Kim:1985ez}
  H.~J.~Kim, L.~J.~Romans and P.~van Nieuwenhuizen,
  ``The Mass Spectrum Of Chiral N=2 D=10 Supergravity On S**5,''
  Phys.\ Rev.\  D {\bf 32} (1985) 389.
  
\bibitem{Randall:1999vf}
  L.~Randall and R.~Sundrum,
  ``An alternative to compactification,''
  Phys.\ Rev.\ Lett.\  {\bf 83} (1999) 4690
  [arXiv:hep-th/9906064].

\bibitem{Langfelder:2006vd}
  P.~Langfelder,
  ``On tunnelling in two-throat warped reheating,''
  JHEP {\bf 0606} (2006) 063
  [arXiv:hep-th/0602296].

\bibitem{Harling:2007jy}
  B.~v.~Harling, A.~Hebecker and T.~Noguchi,
  ``Energy Transfer between Throats from a 10d Perspective,''
  JHEP {\bf 0711} (2007) 042
  [arXiv:0705.3648 [hep-th]].

\bibitem{Kuperstein:2004hy}
  S.~Kuperstein,
  ``Meson spectroscopy from holomorphic probes on the warped deformed
  conifold,''
  JHEP {\bf 0503} (2005) 014
  [arXiv:hep-th/0411097].

\bibitem{Gherghetta:2006yq}
  T.~Gherghetta and J.~Giedt,
  ``Bulk fields in AdS(5) from probe D7 branes,''
  Phys.\ Rev.\  D {\bf 74} (2006) 066007
  [arXiv:hep-th/0605212].

\bibitem{Levi:2005hh}
  T.~S.~Levi and P.~Ouyang,
  ``Mesons and Flavor on the Conifold,''
  Phys.\ Rev.\  D {\bf 76} (2007) 105022
  [arXiv:hep-th/0506021].

\bibitem{Chen:2008jj}
  H.~Y.~Chen, P.~Ouyang and G.~Shiu,
  ``On Supersymmetric D7-branes in the Warped Deformed Conifold,''
  arXiv:0807.2428 [hep-th].

\bibitem{Maldacena:1997re}
  J.~M.~Maldacena,
  ``The large N limit of superconformal field theories and supergravity,''
  Adv.\ Theor.\ Math.\ Phys.\  {\bf 2} (1998) 231
  [Int.\ J.\ Theor.\ Phys.\  {\bf 38} (1999) 1113]
  [arXiv:hep-th/9711200], \\
  O.~Aharony, S.~S.~Gubser, J.~M.~Maldacena, H.~Ooguri and Y.~Oz,
  ``Large N field theories, string theory and gravity,''
  Phys.\ Rept.\  {\bf 323} (2000) 183
  [arXiv:hep-th/9905111].

\bibitem{Klebanov:1999tb}
  I.~R.~Klebanov and E.~Witten,
  ``AdS/CFT correspondence and symmetry breaking,''
  Nucl.\ Phys.\  B {\bf 556} (1999) 89
  [arXiv:hep-th/9905104].

\bibitem{Gubser:1998bc}
  S.~S.~Gubser, I.~R.~Klebanov and A.~M.~Polyakov,
  ``Gauge theory correlators from non-critical string theory,''
  Phys.\ Lett.\  B {\bf 428} (1998) 105
  [arXiv:hep-th/9802109].

\bibitem{Witten:1998zw}
  E.~Witten,
  ``Anti-de Sitter space and holography,''
  Adv.\ Theor.\ Math.\ Phys.\  {\bf 2} (1998) 253
  [arXiv:hep-th/9802150].

\bibitem{PerezVictoria:2001pa}
  M.~Perez-Victoria,
  ``Randall-Sundrum models and the regularized AdS/CFT correspondence,''
  JHEP {\bf 0105} (2001) 064
  [arXiv:hep-th/0105048].

\bibitem{Gherghetta:2006ha}
  T.~Gherghetta,
  ``Warped models and holography,''
  arXiv:hep-ph/0601213.

\bibitem{Batell:2007jv}
  B.~Batell and T.~Gherghetta,
  ``Holographic Mixing Quantified,''
  Phys.\ Rev.\  D {\bf 76} (2007) 045017
  [arXiv:0706.0890 [hep-th]], \\
  B.~Batell and T.~Gherghetta,
  ``Warped Phenomenology in the Holographic Basis,''
  Phys.\ Rev.\  D {\bf 77} (2008) 045002
  [arXiv:0710.1838 [hep-ph]].

\bibitem{Klebanov:1998hh}
  I.~R.~Klebanov and E.~Witten,
  ``Superconformal field theory on threebranes at a Calabi-Yau  singularity,''
  Nucl.\ Phys.\  B {\bf 536} (1998) 199
  [arXiv:hep-th/9807080].
  
\bibitem{Witten:2001ua}
  E.~Witten,
  ``Multi-trace operators, boundary conditions, and AdS/CFT correspondence,''
  arXiv:hep-th/0112258.

\bibitem{Moroz:2009kv}
  S.~Moroz,
  ``Below the Breitenlohner-Freedman bound in the nonrelativistic AdS/ CFT
  correspondence,''
  arXiv:0911.4060 [hep-th];\\
  O.~Antipin and K.~Tuominen,
  ``Resizing the Conformal Window: A beta function Ansatz,''
  arXiv:0909.4879 [hep-ph].

\bibitem{Klebanov:1999rd}
  I.~R.~Klebanov and N.~A.~Nekrasov,
  ``Gravity duals of fractional branes and logarithmic RG flow,''
  Nucl.\ Phys.\  B {\bf 574} (2000) 263
  [arXiv:hep-th/9911096].

\bibitem{Klebanov:2000nc}
  I.~R.~Klebanov and A.~A.~Tseytlin,
  ``Gravity Duals of Supersymmetric SU(N) x SU(N+M) Gauge Theories,''
  Nucl.\ Phys.\  B {\bf 578} (2000) 123
  [arXiv:hep-th/0002159].
  
\bibitem{Denef:2008wq}
  F.~Denef,
  ``Les Houches Lectures on Constructing String Vacua,''
  arXiv:0803.1194 [hep-th].

\bibitem{Klebanov:2002gr}
  I.~R.~Klebanov, P.~Ouyang and E.~Witten,
  ``A gravity dual of the chiral anomaly,''
  Phys.\ Rev.\  D {\bf 65} (2002) 105007
  [arXiv:hep-th/0202056].
  
\bibitem{Aharony:2000pp}
  O.~Aharony,
  ``A note on the holographic interpretation of string theory backgrounds  with
  varying flux,''
  JHEP {\bf 0103} (2001) 012
  [arXiv:hep-th/0101013].

\bibitem{Gubser:2004qj}
  S.~S.~Gubser, C.~P.~Herzog and I.~R.~Klebanov,
  ``Symmetry breaking and axionic strings in the warped deformed conifold,''
  JHEP {\bf 0409} (2004) 036
  [arXiv:hep-th/0405282].

\bibitem{Gubser:2004tf}
  S.~S.~Gubser, C.~P.~Herzog and I.~R.~Klebanov,
  ``Variations on the warped deformed conifold,''
  Comptes Rendus Physique {\bf 5} (2004) 1031
  [arXiv:hep-th/0409186].

\bibitem{Morrison:1998cs}
  D.~R.~Morrison and M.~R.~Plesser,
  ``Non-spherical horizons. I,''
  Adv.\ Theor.\ Math.\ Phys.\  {\bf 3}, 1 (1999)
  [arXiv:hep-th/9810201].

\bibitem{Butti:2004pk}
  A.~Butti, M.~Grana, R.~Minasian, M.~Petrini and A.~Zaffaroni,
  ``The baryonic branch of Klebanov-Strassler solution: A supersymmetric
  family of SU(3) structure backgrounds,''
  JHEP {\bf 0503} (2005) 069
  [arXiv:hep-th/0412187].

\bibitem{Papadopoulos:2000gj}
  G.~Papadopoulos and A.~A.~Tseytlin,
  ``Complex geometry of conifolds and 5-brane wrapped on 2-sphere,''
  Class.\ Quant.\ Grav.\  {\bf 18} (2001) 1333
  [arXiv:hep-th/0012034].

\bibitem{DeWolfe:2004qx}
  O.~DeWolfe, S.~Kachru and H.~L.~Verlinde,
  ``The giant inflaton,''
  JHEP {\bf 0405} (2004) 017
  [arXiv:hep-th/0403123].

\bibitem{Aharony:2005ez}
  O.~Aharony, Y.~E.~Antebi and M.~Berkooz,
  ``Open string moduli in KKLT compactifications,''
  Phys.\ Rev.\  D {\bf 72} (2005) 106009
  [arXiv:hep-th/0508080].

\bibitem{Benna:2007mb}
  M.~K.~Benna, A.~Dymarsky, I.~R.~Klebanov and A.~Solovyov,
  ``On Normal Modes of a Warped Throat,''
  JHEP {\bf 0806} (2008) 070
  [arXiv:0712.4404 [hep-th]].

\bibitem{Dymarsky:2008wd}
  A.~Dymarsky, D.~Melnikov and A.~Solovyov,
  ``I-odd sector of the Klebanov-Strassler theory,''
  JHEP {\bf 0905} (2009) 105
  [arXiv:0810.5666 [hep-th]].

\bibitem{Krasnitz:2000ir}
  M.~Krasnitz,
  ``A two point function in a cascading N = 1 gauge theory from
  supergravity,''
  arXiv:hep-th/0011179.

\bibitem{Caceres:2000qe}
  E.~Caceres and R.~Hernandez,
  ``Glueball masses for the deformed conifold theory,''
  Phys.\ Lett.\  B {\bf 504} (2001) 64
  [arXiv:hep-th/0011204].

\bibitem{Amador:2004pz}
  X.~Amador and E.~Caceres,
  ``Spin two glueball mass and glueball Regge trajectory from supergravity,''
  JHEP {\bf 0411} (2004) 022
  [arXiv:hep-th/0402061].

\bibitem{Noguchi:2005ws}
  T.~Noguchi, M.~Yamaguchi and M.~Yamashita,
  ``Gravitational Kaluza-Klein modes in warped superstring  compactification,''
  Phys.\ Lett.\  B {\bf 636} (2006) 221
  [arXiv:hep-th/0512249].

\bibitem{Berg:2006xy}
  M.~Berg, M.~Haack and W.~Mueck,
  ``Glueballs vs. gluinoballs: Fluctuation spectra in non-AdS/non-CFT,''
  Nucl.\ Phys.\  B {\bf 789} (2008) 1
  [arXiv:hep-th/0612224].

\bibitem{Dymarsky:2007zs}
  A.~Dymarsky and D.~Melnikov,
  ``Gravity Multiplet on KS and BB Backgrounds,''
  JHEP {\bf 0805} (2008) 035
  [arXiv:0710.4517 [hep-th]].

\bibitem{Douglas:2008jx}
  M.~R.~Douglas and G.~Torroba,
  ``Kinetic terms in warped compactifications,''
  arXiv:0805.3700 [hep-th].

\bibitem{Frey:2008xw}
  A.~R.~Frey, G.~Torroba, B.~Underwood and M.~R.~Douglas,
  ``The Universal Kaehler Modulus in Warped Compactifications,''
  JHEP {\bf 0901} (2009) 036
  [arXiv:0810.5768 [hep-th]].

\bibitem{Barnaby:2004kz}
  N.~Barnaby and J.~M.~Cline,
  ``Tachyon defect formation and reheating in brane-antibrane inflation,''
  Int.\ J.\ Mod.\ Phys.\  A {\bf 19} (2004) 5455
  [arXiv:hep-th/0410030] and 
  ``Creating the universe from brane-antibrane annihilation,''
  Phys.\ Rev.\  D {\bf 70} (2004) 023506
  [arXiv:hep-th/0403223].

\bibitem{Baumann:2008kq}
  D.~Baumann, A.~Dymarsky, S.~Kachru, I.~R.~Klebanov and L.~McAllister,
  ``Holographic Systematics of D-brane Inflation,''
  JHEP {\bf 0903} (2009) 093
  [arXiv:0808.2811 [hep-th]].

\bibitem{Baumann:2006th}
  D.~Baumann, A.~Dymarsky, I.~R.~Klebanov, J.~M.~Maldacena, L.~P.~McAllister and A.~Murugan,
  ``On D3-brane potentials in compactifications with fluxes and wrapped
  D-branes,''
  JHEP {\bf 0611} (2006) 031
  [arXiv:hep-th/0607050];\\
  D.~Baumann, A.~Dymarsky, I.~R.~Klebanov, L.~McAllister and P.~J.~Steinhardt,
  ``A Delicate Universe,''
  Phys.\ Rev.\ Lett.\  {\bf 99} (2007) 141601
  [arXiv:0705.3837 [hep-th]];\\
  D.~Baumann, A.~Dymarsky, I.~R.~Klebanov and L.~McAllister,
  ``Towards an Explicit Model of D-brane Inflation,''
  JCAP {\bf 0801} (2008) 024
  [arXiv:0706.0360 [hep-th]].

\bibitem{Burgess:2006cb}
  C.~P.~Burgess, J.~M.~Cline, K.~Dasgupta and H.~Firouzjahi,
  ``Uplifting and inflation with D3 branes,''
  JHEP {\bf 0703} (2007) 027
  [arXiv:hep-th/0610320].

\bibitem{Chen:2009nk}\
  H.~Y.~Chen, L.~Y.~Hung and G.~Shiu,
  ``Inflation on an Open Racetrack,''
  JHEP {\bf 0903} (2009) 083
  [arXiv:0901.0267 [hep-th]].

\bibitem{Delbourgo:2000nq}
  L.~Parker,
  ``Gravitons from anomalous decay,''
  Int.\ J.\ Theor.\ Phys.\  {\bf 28} (1989) 1163;\\
  R.~Delbourgo and D.~s.~Liu,
  ``Electromagnetic and gravitational decay of the Higgs boson,''
  Austral.\ J.\ Phys.\  {\bf 53}, 647 (2001)
  [arXiv:hep-ph/0004156].

\bibitem{Kawasaki:2000en}
  M.~Kawasaki, K.~Kohri and N.~Sugiyama,
  ``MeV-scale reheating temperature and thermalization of neutrino
  background,''
  Phys.\ Rev.\  D {\bf 62}, 023506 (2000)
  [arXiv:astro-ph/0002127];\\
  G.~F.~Giudice, E.~W.~Kolb and A.~Riotto,
  ``Largest temperature of the radiation era and its cosmological
  implications,''
  Phys.\ Rev.\  D {\bf 64}, 023508 (2001)
  [arXiv:hep-ph/0005123];\\
  S.~Hannestad,
  ``What is the lowest possible reheating temperature?,''
  Phys.\ Rev.\  D {\bf 70}, 043506 (2004)
  [arXiv:astro-ph/0403291].

\bibitem{Kallosh:2004yh}
  R.~Kallosh and A.~D.~Linde,
  ``Landscape, the scale of SUSY breaking, and inflation,''
  JHEP {\bf 0412} (2004) 004
  [arXiv:hep-th/0411011].

\bibitem{Conlon:2008cj}
  J.~P.~Conlon, R.~Kallosh, A.~D.~Linde and F.~Quevedo,
  ``Volume Modulus Inflation and the Gravitino Mass Problem,''
  JCAP {\bf 0809} (2008) 011
  [arXiv:0806.0809 [hep-th]].
 
 \bibitem{Kribs:1996ac}
 G.~D.~Kribs and I.~Z.~Rothstein,
 ``Bounds on long-lived relics from diffuse gamma ray observations,''
 Phys.\ Rev.\ D {\bf 55}, 4435 (1997)
 [Erratum-ibid.\ D {\bf 56}, 1822 (1997)]
 [arXiv:hep-ph/9610468].

\bibitem{Falkowski:2009yz}
  A.~Falkowski, J.~Juknevich and J.~Shelton,
  ``Dark Matter Through the Neutrino Portal,''
  arXiv:0908.1790 [hep-ph].


\bibitem{Davoudiasl:1999tf}
  H.~Davoudiasl, J.~L.~Hewett and T.~G.~Rizzo,
  ``Bulk gauge fields in the Randall-Sundrum model,''
  Phys.\ Lett.\  B {\bf 473} (2000) 43
  [arXiv:hep-ph/9911262].

\bibitem{Chung:2000rg}
  D.~J.~H.~Chung, L.~L.~Everett and H.~Davoudiasl,
  ``Experimental probes of the Randall-Sundrum infinite extra dimension,''
  Phys.\ Rev.\  D {\bf 64} (2001) 065002
  [arXiv:hep-ph/0010103].

\bibitem{Abramowitz}
  M.~Abramowitz and I.~A. Stegun, eds., 
  ``Handbook of Mathematical Functions'', \\
  New York USA: Dover Publ., 1965.

\end{thebibliography}
\end{document}